\documentclass[reqno,11pt]{amsart}
\usepackage{graphicx}
\usepackage{amscd}
\usepackage{slashed}
\usepackage{amssymb}
\usepackage{esint}
\usepackage[mathscr]{eucal}
\numberwithin{equation}{section}
\allowdisplaybreaks[1]

\newcommand{\SetFigFont}[3]{}

\title[The Fermionic Projector in Space-Times of Infinite Lifetime]{A Non-Perturbative Construction of the
Fermionic Projector on Globally Hyperbolic Manifolds I -- Space-Times of Finite Lifetime}

\author[F.\ Finster]{Felix Finster}
\address{Fakult\"at f\"ur Mathematik \\ Universit\"at Regensburg \\ D-93040 Regensburg \\ Germany}
\email{finster@ur.de}

\author[M.\ Reintjes]{Moritz Reintjes \\ \\ January 2013}
\address{IMPA - Instituto Nacional de Matem{\'a}tica Pura e Aplicada \\ Rio de Janeiro, Brasil}
\email{moritzreintjes@gmail.com}
\thanks{M.R.\ is supported by the Deutsche Forschungsgemeinschaft (DFG)}

\newtheorem{Def}{Definition}[section]
\newtheorem{Thm}[Def]{Theorem}
\newtheorem{Prp}[Def]{Proposition}
\newtheorem{Lemma}[Def]{Lemma}

\newtheorem{Corollary}[Def]{Corollary}

\newcommand{\Thanks}{\vspace*{.5em} \noindent \thanks}
\newcommand{\beq}{\begin{equation}}
\newcommand{\eeq}{\end{equation}}
\newcommand{\Proof}{\begin{proof}}
\newcommand{\QED}{\end{proof} \noindent}

\newcommand{\la}{\langle}
\newcommand{\ra}{\rangle}
\newcommand{\bra}{\mathopen{<}}
\newcommand{\ket}{\mathclose{>}}
\newcommand{\Sl}{\mathopen{\prec}}
\newcommand{\Sr}{\mathclose{\succ}}

\newcommand{\C}{\mathbb{C}}
\newcommand{\R}{\mathbb{R}}
\newcommand{\1}{\mbox{\rm 1 \hspace{-1.05 em} 1}}

\newcommand{\N}{\mathbb{N}}

\newcommand{\nuslsh}{\slashed{\nu}}
\renewcommand{\H}{\mathscr{H}}

\newcommand{\U}{{\rm{U}}}

\newcommand{\uslsh}{\slashed{u}}

\newcommand{\bep}{\begin{pmatrix}}
\newcommand{\enp}{\end{pmatrix}}

\newcommand{\F}{{\mathscr{F}}}
\newcommand{\Dir}{{\mathcal{D}}}
\newcommand{\D}{{\mathscr{D}}}

\newcommand{\B}{{\mathscr{B}}}
\renewcommand{\O}{{\mathscr{O}}}

\newcommand{\Lin}{\text{\rm{L}}}

\newcommand{\Cisc}{C^\infty_{\text{sc}}}

\DeclareMathOperator{\Tr}{Tr}

\DeclareMathOperator{\supp}{supp}

\newcommand{\Sig}{\mathscr{S}}
\newcommand{\scrM}{\mycal M}
\newcommand{\scrN}{\mycal N}

\DeclareFontFamily{OT1}{rsfso}{}
\DeclareFontShape{OT1}{rsfso}{m}{n}{ <-7> rsfso5 <7-10> rsfso7 <10-> rsfso10}{}
\DeclareMathAlphabet{\mycal}{OT1}{rsfso}{m}{n}

\begin{document}
\maketitle

\begin{abstract}
We give a functional analytic construction of the fermionic projector on a globally
hyperbolic Lorentzian manifold of finite lifetime.
The integral kernel of the fermionic projector is represented by a two-point distribution on the manifold.
By introducing an ultraviolet regularization, we get to the framework of causal fermion systems.
The connection to the ``negative-energy solutions'' of the Dirac equation and to the WKB approximation
is explained and quantified by a detailed analysis of
closed Friedmann-Robertson-Walker universes.
\end{abstract}

\tableofcontents

\section{Introduction}
The fermionic projector was introduced in~\cite{sea} as an operator which gives a splitting of the
solution space of the Dirac equation into two subspaces (see also~\cite[Chapter~2]{PFP} and~\cite{grotz}).
In a static space-time, these subspaces reduce to the spaces of positive and
negative energy which are familiar from the usual Dirac sea construction.
The significance of the fermionic projector lies in the fact that it can be constructed canonically
even in the time-dependent setting. It plays a central role in the fermionic projector approach
to relativistic quantum field theory (see the review article~\cite{srev} and the references therein).

So far, the fermionic projector was only constructed perturbatively in a formal power expansion
in the potentials in the Dirac equation.
In the present paper, we give a {\em{non-perturbative construction}}
of the fermionic projector. To this end, we consider the Dirac equation on a globally hyperbolic
Lorentzian manifold. For technical simplicity, we assume that space-time has finite lifetime.
A space-time of infinite lifetime (like Minkowski space) can be treated with the same
ideas and methods, using the so-called mass oscillation property as an additional technical tool.
Since the mass oscillation property is of independent interest, we decided to work
out the case of an infinite lifetime in a separate paper~\cite{infinite}.

In order to explain the basic difficulty which prevented a non-perturbative treatment so far,
we briefly outline the construction in~\cite{sea} on a non-technical level.
Suppose that we consider the Dirac equation in Minkowski space~$(\scrM, \la .,. \ra)$ in a given external
potential~$\B$,
\[ (i \gamma^j \partial_j + \B - m) \psi = 0 \:. \]
Then the advanced and retarded Green's functions~$s^\vee_m$ and~$s^\wedge_m$ are
solutions of the distributional equations
\[ (i \gamma^j \partial_j + \B - m) \,s^\vee_m(x,y) = \delta^4(x-y) = (i \gamma^j \partial_j + \B - m) \,s^\wedge_m(x,y) \:. \]
They are uniquely defined by the conditions that the distribution~$s^\vee(x,.)$ 
(and~$s^\wedge(x,.)$) should be supported in the causal future (respectively past) of~$x$.
Taking the difference of the advanced and retarded Green's function gives 
a solution of the homogeneous Dirac equation, which we refer to as the 
causal fundamental solution~$k_m$,
\[ k_m(x,y) := \frac1{2\pi i}\left( s^{\vee}(x,y)-s^{\wedge}(x,y) \right) . \]
We also consider~$k_m$ as the integral kernel of a corresponding operator
\[ (k_m(\psi))(x) := \int_\scrM k_m(x,y)\: \psi(y)\: d^4y \:, \]
which acts on the wave functions in space-time.
Here the integral merely is a notation to indicate a distribution acting on a test function
(thus~$k_m(\psi) = k_m(., \psi)$ is the distribution obtained by evaluating the
second argument of the bi-distribution~$k_m(.,.)$ with~$\psi$).
Formally, the fermionic projector is
obtained by taking the absolute value of this operator,
\beq \label{kabs}
p_m \overset{\text{formally}}{:=} |k_m| \:,
\eeq
and by forming the combination
\[ P(x,y) := \frac{1}{2} \left( p_m(x,y) - k_m(x,y) \right) \]
(for the rescaling procedure needed to obtain the proper normalization see~\cite{grotz}).
The basic difficulty is related to the fact that taking the absolute value of $k_m$ in a rigorous way requires spectral methods in Hilbert spaces. But the operator~$k_m$ acts on the wave functions
in space-time, which do not form a Hilbert space.
More specifically, $k_m$ is symmetric with respect to the Lorentz invariant inner product on the wave functions
\beq \label{stipMin}
\bra \psi|\phi \ket = \int_\scrM \overline{\psi(x)} \phi(x) \: d^4x
\eeq
(where~$\overline{\psi} \equiv \psi^\dagger \gamma^0$ is the so-called adjoint spinor;
we here restrict attention to square integrable wave functions).
But as~\eqref{stipMin} is not positive definite, the corresponding function space
merely is a Krein space. There is a spectral theorem in Krein spaces (see
for example~\cite{bognar, langer}), but this theorem only applies to so-called
definitizable operators. The operator~$k_m$, however, is not known to be definitizable,
making it impossible to apply spectral methods in indefinite inner product spaces.
The methods in~\cite{sea} give a mathematical meaning to the absolute value in~\eqref{kabs} 
in a perturbation expansion, leading to the so-called causal perturbation theory.
But a non-perturbative treatment seemed out of reach.

We now outline our method for bypassing the above difficulty, again for an external
potential in flat space-time. One ingredient is to work instead of the space of
wave functions with the solution space of the Dirac equation. This solution space has a natural Lorentz
invariant scalar product
\beq \label{sprodMin}
( \psi | \phi) := \int_{\R^3} (\overline{\psi} \gamma^0 \phi)(t, \vec{x})\: d^3x \:,
\eeq
giving rise to a Hilbert space~$\H_m$. Our starting point is the observation
(see~\cite[Proposition~2.2]{dimock3}) that the operator~$k_m$
relates the scalar product~\eqref{sprodMin} to the space-time inner product~\eqref{stipMin} by
\beq \label{rel1}
(\psi \,|\, k_m \,\phi) = \bra \psi | \phi \ket
\eeq
(valid if~$\psi$ is a solution of the Dirac equation; see Proposition~\ref{prpdual} below).
On the other hand, we can express the bilinear form~$\bra .|. \ket$ in terms of
the scalar product using a signature operator~$\Sig$,
\beq \label{rel2}
\bra \psi | \phi \ket = ( \psi | \Sig \phi )
\eeq
(valid if~$\psi$ and~$\phi$ are solutions of the Dirac equation; see equation~\eqref{Sdef} below).
The operator~$\Sig$ will turn out to be a bounded symmetric operator on the Hilbert space~$(\H_m, (.|.))$.
Comparing~\eqref{rel1} with~\eqref{rel2}, we find that on solutions of the Dirac equation,
the operator~$k_m$ can be identified with the operator~$\Sig$. This makes it possible to use spectral theory in Hilbert spaces to define the absolute value in~\eqref{kabs}.

In Section~\ref{secFA}, we will make this construction mathematically precise
in the setting of a globally hyperbolic space-time of finite lifetime.
We point out that all our constructions are manifestly covariant. They do not depend on the
choice of a foliation of the manifold. It makes no difference whether the Cauchy surfaces are
compact or non-compact. We do not need to make any assumptions on
the asymptotic behavior of the metric at infinity.

In Section~\ref{sec4}, it is explained how the fermionic projector gives rise to examples
of causal fermion systems as defined in~\cite[Section~1]{rrev}.

Our construction of the fermionic projector gives a splitting of the solution space of the
Dirac equation into two subspaces. For the physical interpretation, it is important
to understand how these subspaces relate to the usual concept of solutions of positive and negative energy.
To this end, we analyze the fermionic projector in a closed Friedmann-Robertson-Walker universe.
This has the advantage that the Dirac equation reduces to an ODE in time, which can
be analyzed in detail. In particular, the concept of ``solutions of negative energy'' 
(which for clarity we mostly refer to as ``solutions of negative frequency'') can be
made precise by a specific WKB approximation as worked out in~\cite{moritz}.
In Section~\ref{secFRW}, it is shown that our definition of the fermionic projector
agrees with the concept of ``all solutions of negative frequency,'' provided that the metric
is ``nearly constant'' on the Compton scale as
quantified in Theorem~\ref{thmPWKB} and Theorem~\ref{thmPmP}.
It is remarkable that, in contrast to a Gr\"onwall estimate,
our error estimates do not involve a time integral of the error term. This means that small
local errors of the WKB approximation do not ``add up'' to give a big error after a long time.
Moreover, our estimates also apply near the big bang and big crunch singularities.
Keeping these facts in mind, our estimates show that for our physical universe, the fermionic projector
coincides with very high precision with the usual concept of the Dirac sea being composed
of all negative-frequency solutions of the Dirac equation. This gives a rigorous justification
of the physical concepts behind the fermionic projector approach.

In Section~\ref{seccounter}, we analyze what happens if the metric changes substantially
on the Compton scale. To this end, we consider a closed Friedmann-Robertson-Walker universe
with a scale function~$R(\tau)$ being piecewise constant.
Then, at the times when~$R$ is discontinuous, the frequencies of the solutions change. As a consequence, the concept of positive or negative frequency becomes meaningless.
In this situation, our constructions still apply, giving a well-defined fermionic projector.
This fermionic projector consists of a mixture of positive and negative frequencies.
Moreover, as we explain in an explicit example where~$\Sig=0$, the fermionic projector may depend
sensitively on the detailed geometry of space-time.

\section{Preliminaries}
Let~$(\scrM, g)$ be a smooth, globally hyperbolic Lorentzian manifold of dimension~$k \geq 2$.
For the signature of the metric we use the convention~$(+ ,-, \ldots, -)$.
As proven in~\cite{bernal+sanchez}, $\scrM$ admits a smooth foliation~$(\scrN_t)_{t \in \R}$
by Cauchy hypersurfaces. Thus~$\scrM$ is topologically the product of~$\R$ with a $k-1$-dimensional manifold.
In the case~$k=4$ of a four-dimensional space-time, this implies that~$\scrM$ is spin
(for details see~\cite{baum, lawson+michelsohn}).
For a general space-time dimension we need to impose that~$\scrM$ is spin.
We let~$S\scrM$ be the spinor bundle on~$\scrM$ and denote the smooth sections of the spinor bundle
by~$C^\infty(\scrM, S\scrM)$. Similarly, $C^\infty_0(\scrM, S\scrM)$ denotes the smooth sections with compact support.
The sections of the spinor bundle are also referred to as wave functions.
The fibres~$S_x\scrM$ are endowed with an inner product of signature~$(n,n)$
with~$n=2^{[k/2]-1}$ (where~$[\cdot]$ is the Gau{\ss} bracket; for details see
again~\cite{baum, lawson+michelsohn}),
which we denote by~$\Sl .|. \Sr_x$. The Lorentzian metric induces a Levi-Civita connection
and a spin connection, which we both denote by~$\nabla$.
Every vector of the tangent space acts on the corresponding spinor space by Clifford multiplication.
Clifford multiplication is related to the Lorentzian metric via the anti-commutation relations.
Denoting the mapping from the
tangent space to the linear operators on the spinor space by~$\gamma$, we thus have
\[ \gamma \::\: T_x\scrM \rightarrow \Lin(S_x\scrM) \qquad
\text{with} \qquad \gamma(u) \,\gamma(v) + \gamma(v) \,\gamma(u) = 2 \, g(u,v)\,\1_{S_x(\scrM)} \:. \]
We also write Clifford multiplication in components with the Dirac matrices~$\gamma^j$
and use the short notation with the Feynman dagger, $\gamma(u) \equiv u^j \gamma_j \equiv \uslsh$.
The connections, inner products and Clifford multiplication satisfy Leibniz rules and compatibility
conditions; we refer to~\cite{baum, lawson+michelsohn} for details.
Combining the spin connection with Clifford multiplication gives the geometric Dirac operator~$\Dir = i \gamma^j
\nabla_j$. In order to include the situation when an external 
potential is present, we add a multiplication operator~$\B(x) \in \Lin(S_x\scrM)$, which we assume to
be smooth and symmetric with respect to the spin scalar product,
\beq \label{Bdef}
\B \in C^\infty(\scrM, \Lin(S\scrM)) \qquad \text{with} \qquad
\Sl \B \phi | \psi \Sr_x = \Sl \phi | \B \psi \Sr_x \quad \forall \phi, \psi \in S_x\scrM\:.
\eeq
We then introduce the Dirac operator by
\beq \label{Dirdef}
\Dir := i \gamma^j \nabla_j + \B \::\: C^\infty(\scrM, S\scrM) \rightarrow C^\infty(\scrM, S\scrM)\:.
\eeq
For a given real parameter~$m$ (the ``rest mass''), the Dirac equation reads
\beq \label{dirac}
(\Dir - m) \,\psi_m = 0 \:.
\eeq
For clarity, solutions of the Dirac equation always carry a subscript~$m$.
We point out that throughout this paper, the case~$m=0$ of a massless
field is allowed.

In the Cauchy problem, one seeks for a solution of the Dirac equation with
initial data~$\psi_\scrN$ prescribed on a given Cauchy surface~$\scrN$. Thus in the smooth setting,
\beq \label{cauchy}
(\D - m) \,\psi_m = 0 \:,\qquad \psi|_{\scrN} = \psi_\scrN \in C^\infty(\scrN, S\scrM) \:.
\eeq
This Cauchy problem has a unique solution~$\psi_m \in C^\infty(\scrM, S\scrM)$.
This can be seen either by considering energy estimates for symmetric hyperbolic systems
(see for example~\cite{john}) or alternatively by constructing the Green's kernel (see for
example~\cite{baer+ginoux}). These methods also show that the Dirac equation is causal,
meaning that the solution of the Cauchy problem only depends on the initial data in the causal
past or future. In particular, if~$\psi_\scrN$ has compact support, the solution~$\psi_m$ will also have compact
support on any other Cauchy hypersurface. This leads us to consider solutions~$\psi_m$
in the class~$\Cisc(\scrM, S\scrM)$ of smooth sections with spatially compact support. On solutions in this class,
one introduces the scalar product~$(.|.)_\scrN$ by\footnote{The factor~$2 \pi$ might seem
unconventional. This convention was first adopted in~\cite{rrev}. It will simplify
many formulas in this paper.}
\beq \label{print}
(\psi_m | \phi_m)_\scrN = 2 \pi \int_\scrN \Sl \psi_m | \nuslsh \phi_m \Sr_x\: d\mu_\scrN(x) \:,
\eeq
where~$\nuslsh$ denotes Clifford multiplication by the future-directed normal~$\nu$
(we always adopt the convention that the inner product~$\Sl . | \nuslsh . \Sr_x$ is
{\em{positive}} definite).
This scalar product does not depend on the choice of the Cauchy surface~$\scrN$.
To see this, we let~$\scrN'$ be another Cauchy surface and~$\Omega$ the space-time
region enclosed by~$\scrN$ and~$\scrN'$. Using the symmetry property
in~\eqref{Bdef} together with~\eqref{Dirdef} and~\eqref{dirac}, we obtain
\beq \begin{split} \label{divfree}
i \nabla_j \Sl \psi_m | \gamma^j \phi_m \Sr_x &=
\Sl (-i \nabla_j) \psi_m | \gamma^j \phi_m \Sr_x
+\Sl \psi_m | (i \gamma^j \nabla_j) \phi_m \Sr_x \\
&= -\Sl \Dir \psi_m | \phi_m \Sr_x
+\Sl \psi_m | \Dir \phi_m \Sr_x = 0\:,
\end{split}
\eeq
showing that the vector field~$\Sl \psi_m | \gamma^j \phi_m \Sr_x$ is divergence-free
(``current conservation''). Integrating over~$\Omega$ and applying the Gau{\ss} divergence
theorem, we find that~$(\psi_m | \phi_m)_\scrN = (\psi_m | \phi_m)_{\scrN'}$.
In view of the independence of the choice of the Cauchy surface, we simply denote
the scalar product~\eqref{print} by~$( .|. )$. Forming the completion, we obtain
the Hilbert space~$(\H_m, (.|.))$. It consists of all weak solutions of the Dirac equation~\eqref{dirac}
which are square integrable over any Cauchy surface.

The {\em{retarded}} and {\em{advanced Green's operators}}~$s_m^\wedge$ and~$s_m^\vee$ are
linear mappings (see for example~\cite{dimock3, baer+ginoux})
\[ s_m^\wedge, s_m^\vee \::\: C^\infty_0(\scrM, S\scrM) \rightarrow \Cisc(\scrM, S\scrM)\:. \]
They satisfy the defining equation of the Green's operator
\beq \label{Greendef}
(\Dir - m) \left( s_m^{\wedge, \vee} \phi \right) = \phi \:.
\eeq
Moreover, they are uniquely determined by the condition that the support of~$s_m^\wedge \phi$
(or~$s_m^\vee \phi$) lies in the future (respectively the past) of~$\supp \phi$.
The {\em{causal fundamental solution}}~$k_m$ is introduced by
\beq \label{kmdef}
k_m := \frac{1}{2 \pi i} \left( s_m^\vee - s_m^\wedge \right) \::\: C^\infty_0(\scrM, S\scrM) \rightarrow \Cisc(\scrM, S\scrM)
\cap \H_m \:.
\eeq
Note that it maps to solutions of the Dirac equation. Moreover,
the distribution~$k_m(x,y)$ can be used to construct an explicit solution of the Cauchy problem, as we recall
in the next lemma. We only sketch the proof, because in Lemma~\ref{lemma44} an independent proof will
be given.
\begin{Lemma} \label{lemma21} The solution of the Cauchy problem~\eqref{cauchy}
has the representation
\[ \psi_m(x) = 2 \pi \int_\scrN k_m(x,y)\, \nuslsh\, \psi_\scrN(y)\: d\mu_\scrN(y)\:, \]
where~$k_m(x,y)$ is the integral kernel of the operator~$k_m$, i.e.
\beq \label{kmkernel}
(k_m \phi)(x) = \int_\scrM k_m(x,y)\, \phi(y)\: d\mu_\scrM(y)
\eeq
(here again the integrals are a notation for a distribution acting on a test function).
\end{Lemma}
\Proof[Sketch of the Proof.] For the proof that~$k_m$ can be represented with an integral
kernel~\eqref{kmkernel} and for analytic details on~$k_m(x,y)$ we refer to~\cite{baer+ginoux}.
In order to prove~\eqref{cauchy}, it suffices to consider a point~$x$ in the future of~$\scrN$,
in which case~\eqref{cauchy} simplifies in view of~\eqref{kmdef} to
\[ \psi_m(x) = i \int_\scrN s^\wedge_m(x,y)\, \nuslsh(y)\, \psi_\scrN(y)\: d\mu_\scrN(y)\:. \]
This identity is derived as follows: We let~$\eta \in C^\infty(\scrM)$ be a function which is identically equal
to one at~$x$ and on~$\scrN$, but such that the function~$\eta \psi_m$ has compact support
(for example, in a foliation~$(\scrN_t)_{t \in \R}$ one can take~$\eta=\chi(t)$ with~$\chi \in C^\infty_0(\R)$).
Then, using~\eqref{Greendef},
\beq \label{seta}
\psi_m(x) = (\eta \psi_m)(x) \overset{(*)}{=} s^\wedge_m \big( (\D - m) (\eta \psi_m) \big)
= s^\wedge_m \big( i \gamma^j (\partial_j \eta)\, \psi_m) \big) \:,
\eeq
where we used~\eqref{Greendef} and the fact that~$\psi_m$ is a solution of the Dirac equation.
In~($*$) we used the identity
\[ \psi = s^\wedge \big( (\D-m) \psi \big) \qquad \text{for~$\psi \in C^\infty_0(\scrM, S\scrM)$}\:, \]
which follows from the uniqueness of the solution of the Cauchy problem, noting
that the function~$\psi - s^\wedge((\D-m) \psi)$ satisfies the Dirac equation and
vanishes in the past of the support of~$\psi$.
To conclude the proof, as the function~$\eta$ in~\eqref{seta} we choose a
sequence~$\eta_\ell$ which converges in the distributional sense to the
function which in the future and past of~$\scrN$ is equal to one and zero, respectively.
\QED

\section{Functional Analytic Construction of the Fermionic Projector} \label{secFA}
\subsection{The Space-Time Inner Product as a Dual Pairing}
On the wave functions, one can introduce the Lorentz invariant inner product
\beq \label{stip}
\bra \psi|\phi \ket := \int_\scrM \Sl \psi | \phi \Sr_x\: d\mu_\scrM\:.
\eeq
In order to ensure that the space-time integral is finite, we assume that one factor
has compact support. In particular, we can regard~$\bra .|. \ket$ as the dual pairing
\[ \bra .|. \ket \::\: \H_m \times C^\infty_0(\scrM, S\scrM) \rightarrow \C\:. \]
The next proposition shows that the causal fundamental solution
is the signature operator of this dual pairing.

\begin{Prp} \label{prpdual}
For any~$\psi_m \in \H_m$ and~$\phi \in C^\infty_0(\scrM, S\scrM)$,
\beq \label{pairing}
(\psi_m \,|\, k_m \,\phi) = \bra \psi_m | \phi \ket \:.
\eeq
\end{Prp}
\Proof We first give the proof under the additional assumption that~$\psi_m \in \Cisc(\scrM, S\scrM)$.
We choose Cauchy surfaces~$\scrN_+$ and~$\scrN_-$ lying in the future and past of~$\supp \phi$,
respectively. Let~$\Omega$ be the space-time region between these two Cauchy surfaces, i.e.\
$\partial \Omega = \scrN_+ \cup \scrN_-$. Then, according to~\eqref{kmdef},
\begin{align*}
(\psi_m \,|\, k_m \,\phi) &= (\psi_m \,|\, k_m \,\phi)_{\scrN_+}
= \frac{i}{2 \pi} \:(\psi_m \,|\, s_m^\wedge \,\phi)_{\scrN_+} \\
&= \frac{i}{2 \pi} \Big[ (\psi_m \,|\, s_m^\wedge \,\phi)_{\scrN_+} - (\psi_m \,|\, s_m^\wedge \,\phi)_{\scrN_-} \Big] \\
&= i \int_\Omega \nabla_j \Sl \psi_m \,|\, \gamma^j s_m^\wedge \phi \Sr_x\: d\mu(x)\:,
\end{align*}
where in the last line we applied the Gau{\ss} divergence theorem and used~\eqref{print}.
Using that~$\psi_m$ satisfies the Dirac equation, a calculation similar to~\eqref{divfree} yields
\[ (\psi_m \,|\, k_m \,\phi) = \int_\Omega \Sl \psi_m \,|\, (\Dir - m) \,s_m^\wedge \phi \Sr_x\: d\mu(x)
\overset{\eqref{Greendef}}{=} \int_\Omega \Sl \psi_m | \phi \Sr_x\: d\mu(x)\:. \]
As~$\phi$ is supported in~$\Omega$, we can extend the last integration to all of~$\scrM$, giving
the result.

In order to extend the result to general~$\psi_m \in \H_m$, we use the following approximation
argument. Let~$\psi_m^{(n)} \in \H_m \cap \Cisc(\scrM, S\scrM)$ be a sequence which converges in~$\H_m$
to~$\psi_m$. Then obviously~$(\psi^{(n)}_m \,|\, k_m \,\phi) \rightarrow (\psi_m \,|\, k_m \,\phi)$.
In order to show that the right side of~\eqref{pairing} also converges, it suffices to prove
that~$\psi_m^{(n)}$ converges in~$L^2_\text{loc}(\scrM, S\scrM)$ to~$\psi_m$.
Thus let~$K \subset \scrM$ be a
compact set contained in the domain of a chart~$(x, U)$. Using Fubini's theorem, we obtain 
for any~$\psi \in \H_m \cap \Cisc(\scrM, S\scrM)$ the estimate
\[ \int_K \Sl \psi | \nuslsh \psi \Sr d\mu_\scrM = \int dx^0 \int \Sl \psi | \nuslsh \psi \Sr \sqrt{|g|} \:d^3x
\leq C(K) \,(\psi | \psi) \:. \]
Applying this estimate to the functions~$\psi = \psi^{(n)}_m - \psi^{(n')}_m$, we see
that~$\psi^{(n)}_m$ converges in~$L^2(K, S\scrM)$ to a function~$\tilde{\psi}$.
This implies that~$\psi^{(n)}_m$ converges to~$\tilde{\psi}$ pointwise almost everywhere
(with respect to the measure~$d\mu_\scrM$).
Moreover, the convergence of~$\psi^{(n)}_m$ in~$\H_m$ to~$\psi_m$ implies that the
restriction of~$\psi^{(n)}_m$ to any Cauchy surface~$\scrN$ converges to~$\psi_m|_\scrN$ 
pointwise almost everywhere (with respect to the measure~$d\mu_\scrN$).
It follows that~$\tilde{\psi} = \psi_m|_K$, concluding the proof.
\QED

\begin{Corollary} The operator~$k_m$, \eqref{kmdef},
is symmetric with respect to the inner product~\eqref{stip}.
\end{Corollary}
\Proof Using Proposition~\ref{prpdual}, we obtain for all~$\phi, \psi \in C^\infty_0(\scrM, S\scrM)$,
\[ \bra k_m \phi \,|\, \psi \ket = (k_m \phi \,|\, k_m \psi) = \bra \phi \,|\, k_m \psi \ket \:, \]
concluding the proof.
\QED

\subsection{Space-Times of Finite Lifetime}
For the construction of the fermionic projector, we need to assume that space-time has the
following property.
\begin{Def} \label{defmfinite}
A globally hyperbolic manifold~$(\scrM,g)$ is said to be {\bf{{\em{m}}-finite}} if
there is a constant~$c>0$ such that for
all~$\phi_m, \psi_m \in \H_m \cap \Cisc(\scrM, S\scrM)$, the
function~$\Sl \phi_m | \psi_m \Sr_x$  is integrable on~$\scrM$ and
\beq \label{stbound}
|\bra \phi_m | \psi_m \ket| \leq c \:\|\phi_m\|\: \|\psi_m\|
\eeq
(where~$\| . \| = (.|.)^\frac{1}{2}$ is the norm on~$\H_m$).
\end{Def} \noindent
Before going on, let us briefly discuss which manifolds are $m$-finite.
\begin{Def} \label{deffinitelife}
A globally hyperbolic manifold~$(\scrM, g)$ has {\bf{finite lifetime}}
if it admits a foliation~$(\scrN_t)_{t \in (t_0, t_1)}$ by Cauchy surfaces with a bounded
time function~$t$ such that the function~$\la \nu, \partial_t \ra$ is bounded on~$\scrM$ (where~$\nu$ denotes
the future-directed normal on~$\scrN_t$ and~$\la \nu, \partial_t \ra \equiv g(\nu, \partial_t)$).
\end{Def}

\begin{Prp} Every globally hyperbolic manifold of finite lifetime is $m$-finite.
\end{Prp}
\Proof Let~$\phi_m, \psi_m \in \Cisc(\scrM, S\scrM)$ be solutions of the Dirac equation~\eqref{dirac}.
Applying Fubini's theorem and decomposing the volume measure, we obtain
\[ \bra \phi_m | \psi_m \ket = \int_\scrM \Sl \phi_m | \psi_m \Sr(x)\: d\mu_\scrM(x) \\
=\int_{t_0}^{t_1} \int_{\scrN_t} \Sl \phi_m | \psi_m \Sr\, \la \nu, \partial_t \ra \,dt \,d\mu_{\scrN_t} \]
and thus
\[ \big| \bra \phi_m | \psi_m \ket \big| \leq \sup_\scrM \la \nu, \partial_t \ra
\int_{t_0}^{t_1} dt \int_{\scrN_t} |\Sl \phi_m | \psi_m \Sr|\,d\mu_{\scrN_t} \:. \]
Estimating the spatial integral by
\[ \int_{\scrN_t} |\Sl \phi_m | \psi_m \Sr|\,d\mu_{\scrN_t} \leq
\int_{\scrN_t} \sqrt{\Sl \phi_m | \nuslsh \phi_m \Sr}\:
\sqrt{\Sl \psi_m | \nuslsh \psi_m \Sr}\:
d\mu_{\scrN_t} \leq \|\phi_m\| \: \|\psi_m\| \:, \]
we conclude that
\[ \big| \bra \phi_m | \psi_m \ket \big| \leq (t_1-t_0)\: \sup_\scrM \la \nu, \partial_t\ra \: \| \phi_m \|\:
\|\psi_m\|\:. \]
A denseness argument gives the result.
\QED

\begin{Prp} On a globally hyperbolic manifold of finite lifetime, there is a constant~$C<\infty$
such that the arc length of every smooth timelike curve is at most~$C$.
\end{Prp}
\Proof Let~$\gamma$ be a timelike geodesic. Possibly after extending it, we can parametrize
it by the time function~$t \in (t_0, t_1)$ of our foliation. Then the vector field~$\dot{\gamma} - \partial_t$ is
tangential to~$\scrN_t$. Hence we can estimate the length of the geodesic by
\[ L(\gamma) = \int_{t_0}^{t_1} \sqrt{\la \dot{\gamma}, \dot{\gamma} \ra}\, dt
\leq \int_{t_0}^{t_1} \sqrt{\la \dot{\gamma}, \nu \ra \la \nu, \dot{\gamma} \ra}\, dt
= \int_{t_0}^{t_1} \la \nu, \partial_t \ra\, dt \leq (t_1-t_0) \sup_\scrM \la \nu, \partial_t \ra \:. \]
This concludes the proof.
\QED
We do not know whether an upper bound on the length of timelike geodesics 
already implies that the space-time has finite lifetime in the sense of Definition~\ref{deffinitelife}.
Moreover, we do not expect that every $m$-finite manifold has finite lifetime.
Unfortunately, entering the study of these questions goes beyond the scope of the present paper.

\subsection{The Fermionic Signature Operator and the Fermionic Projector}
Let us assume that~$(\scrM,g)$ is $m$-finite. Then the space-time inner product can be extended by
continuity to a bilinear form
\[ \bra .|. \ket \::\: \H_m \times \H_m \rightarrow \C\:. \]
Moreover, applying the Riesz representation theorem, we can
uniquely represent this inner product with a signature operator~$\Sig$,
\beq \label{Sdef}
\Sig \::\: \H_m \rightarrow \H_m \qquad \text{with} \qquad
\bra \phi_m | \psi_m \ket = ( \phi_m \,|\, \Sig\, \psi_m) \:.
\eeq
We refer to~$\Sig$ as the {\bf{fermionic signature operator}}.
It is obviously a symmetric operator.
Moreover, it is bounded according to~\eqref{stbound}.
We conclude that it is self-adjoint. The spectral theorem gives the
spectral decomposition
\[ \Sig = \int_{\sigma(\Sig)} \lambda\: dE_\lambda \:, \]
where~$E_\lambda$ is the spectral measure (see for example~\cite{reed+simon}).
The spectral measure gives rise to the spectral calculus
\[ f(\Sig) = \int_{\sigma(\Sig)} f(\lambda)\: dE_\lambda \::\: \H_m \rightarrow \H_m \:, \]
where~$f$ is a bounded Borel function on~$\sigma(\Sig) \subset \R$.

The spectral calculus for the fermionic signature operator is very useful because it gives rise to
a corresponding spectral calculus for the operator~$k_m$, as we now explain.
Multiplying $k_m$ from the left by~$f(\Sig)$ with a bounded Borel function~$f$
gives an operator
\[ f(\Sig)\, k_m \::\: C^\infty_0(\scrM, S\scrM) \rightarrow \H_m \:. \]
This operator is again symmetric with respect to~$\bra .|. \ket$, because
for any~$\phi, \psi \in C^\infty_0(\scrM, S\scrM)$,
\beq \label{symm} \begin{split}
\bra f(\Sig)\, k_m \, \phi \,|\, \psi \ket &= (f(\Sig)\, k_m \phi \,|\, k_m \psi ) \\
&= (k_m \phi \,|\, f(\Sig)\, k_m \psi ) = \bra \phi \,|\,  f(\Sig)\, k_m \, \psi \ket \:,
\end{split} \eeq
where in the first and last equality we applied Proposition~\ref{prpdual}.
In order to make sense of products of such operators, we can consider the
inner product~$\bra f(\Sig) k_m \phi | g(\Sig) k_m \psi \ket$
(where~$f,g$ are bounded Borel functions).
Combining~\eqref{Sdef} with the spectral calculus for~$\Sig$ and Proposition~\ref{prpdual}, we obtain
\beq \label{square} \begin{split}
\bra f(\Sig) \,k_m\, \phi \,|\, g(\Sig) \,k_m \,\psi \ket
&= ( f(\Sig) \,k_m\, \phi \,|\, \Sig\, g(\Sig) \,k_m \,\psi) \\
&= ( k_m\, \phi \,|\, (f g)(\Sig)\:\Sig \,k_m \,\psi)
= \bra \phi \,|\, (f g)(\Sig)\:\Sig \,k_m \,\psi \ket\:.
\end{split} \eeq
In view of~\eqref{symm}, this identity can be written in the suggestive form
\beq \label{prodsugg}
\left( f(\Sig) k_m \right) \left( g(\Sig) k_m \right) \overset{\text{formally}}{=} (f g)(\Sig)\:\Sig \:k_m \:.
\eeq
Note that this last equation makes no direct mathematical sense because
the image of the operator~$g(\Sig) k_m$ does not lie in the domain of~$k_m$, making
it impossible to take the product. However, with~\eqref{symm} and~\eqref{square}
we have given this product a precise mathematical meaning.

We now use this procedure to construct the fermionic projector.
\begin{Def} \label{ferm_proj_Def}
Assume that the globally hyperbolic manifold~$(\scrM,g)$ is $m$-finite (see Definition~\ref{defmfinite}).
Then the operators~$P_\pm \::\: C^\infty_0(\scrM, S\scrM) \rightarrow \H_m$ are defined by
\beq \label{Ppmdef}
P_+ = \chi_{[0, \infty)}(\Sig)\, k_m \qquad \text{and} \qquad P_- = -\chi_{(-\infty, 0)}(\Sig)\, k_m
\eeq
(where~$\chi$ denotes the characteristic function).
The {\bf{fermionic projector}} $P$ is defined by~$P=P_-$.
\end{Def}

\begin{Prp} \label{prpPpm} 
For all~$\phi, \psi \in C^\infty_0(\scrM, S\scrM)$, the operators~$P_\pm$ have the following properties:
\begin{align}
\bra P_\pm \,\phi \,|\, \psi \ket &= \bra \phi \,|\, P_\pm\, \psi \ket &&\hspace*{-2cm} \text{(symmetry)} 
\label{symmetry} \\
\bra P_+ \,\phi \,|\, P_-\, \psi \ket &= 0 &&\hspace*{-2cm} \text{(orthogonality)} \\
\bra P_\pm \,\phi \,|\, P_\pm \,\psi \ket &= \bra \phi \,|\, |\Sig| P_\pm \,\psi \ket &&\hspace*{-2cm}
\text{(normalization)} \:. \label{normalize}
\end{align}
Moreover, the image of~$P_\pm$ is the positive respectively negative
spectral subspace of~$\Sig$, meaning that
\[ \overline{P_+(C^\infty_0(\scrM, S\scrM))} = E_{(0, \infty)}(\H_m) \:,\qquad
\overline{P_-(C^\infty_0(\scrM, S\scrM))} = E_{(-\infty, 0)}(\H_m) \:. \]
\end{Prp}
\Proof This follows immediately from~\eqref{symm}, \eqref{square} and the functional
calculus for self-adjoint operators in Hilbert spaces.
\QED

We finally explain the normalization property~\eqref{normalize}. We first point out that, due to the
factor~$|\Sig|$ on the right of~\eqref{normalize}, the fermionic projector is not idempotent
and is thus {\em{not}} a projection operator.
The projection property could have been arranged by modifying~\eqref{Ppmdef} to
\[ P = -\chi_{(-\infty, 0)}(\Sig)\, |\Sig|^{-1}\, k_m \:. \]
However, we prefer the definition~\eqref{Ppmdef} and the normalization~\eqref{normalize}.
This normalization can be understood by working with a spatial normalization integral, as we now explain.
In view of Lemma~\ref{lemma21}, we can introduce an operator~$\Pi$ by
\beq \label{Pidef}
\Pi \::\: \H_m \rightarrow \H_m \:, \qquad
(\Pi \,\psi_m)(x) = -2 \pi \int_\scrN P(x,y)\, \nuslsh\, (\psi_m)|_\scrN(y)\: d\mu_\scrN(y)  \:,
\eeq
where~$\scrN$ is any Cauchy surface.
\begin{Prp} \label{prpspatnorm}
The operator~$\Pi$ is a projection operator on~$\H_m$.
\end{Prp}
\Proof Combining Lemma~\ref{lemma21} with~\eqref{Ppmdef}, we find that~$\Pi$
coincides with the operator~$\chi_{(-\infty, 0)}(\Sig)$, which is obviously a projection operator.
\QED
Since~\eqref{Pidef} involves a spatial integral, we also refer to~$P$ as the fermionic projector
with {\bf{spatial normalization}}.
We remark that an alternative method for normalizing the fermionic projector is to
work with a $\delta$-normalization in the mass parameter
(for details see~\cite[eqns~(3.19)-(3.21)]{sea} or~\cite{grotz}).
However, this so-called {\em{mass normalization}} can only be used in space-times of infinite lifetime.
A detailed comparison of the spatial normalization and the mass normalization
is given in~\cite{norm}.

\subsection{Explicit Formulas in a Foliation}
It is instructive to supplement the previous abstract constructions by explicit
formulas in a foliation. We always work with the following particularly convenient class of foliations.
As shown in~\cite{bernal+sanchez2, mueller+sanchez}, there are foliations~$(\scrN_t)_{t \in \R}$
by Cauchy surfaces where the gradient of the time function is orthogonal to the leaves and the lapse
function is bounded, i.e.\
\beq \label{folio}
g = \beta^2 \,dt^2 - g_{\scrN_t} \qquad \text{with} \qquad 0 < \beta \leq 1\:,
\eeq
where~$g_{\scrN_t}$ is the induced Riemannian metric on~$\scrN_t$, and the lapse function~$\beta$
is a smooth function on~$\scrM$.
We remark that in space-times of finite life time (see Definition~\ref{deffinitelife}), the time parameter~$t$
could be chosen on a bounded interval. In this case, for convenience we
prefer to parametrize~$t$ on all of~$\R$, such that~$\lim\limits_{t \rightarrow \pm \infty} \beta = 0$.
We denote space-time points by~$(t,x)$ with~$t \in \R$ and~$x \in \scrN_t$.
Moreover, we denote the scalar product~\eqref{print} for~$\scrN=\scrN_t$ by~$(.|.)_t$, and the
corresponding Hilbert space by~$(\H_t, (.|.)_t)$. Solving the Cauchy problem with initial
data on~$\scrN_t$ and evaluating the solution at another time~$t'$ gives rise to a unitary
time evolution operator
\[ U^{t', t} \::\: \H_t \rightarrow \H_{t'}\:. \]
Clearly, the unitary time evolution operators are a representation of the group~$(\R, +)$.
The time evolution also gives rise to the unitary mapping
\[ \iota_m \::\: \H_t \rightarrow \H_m \:,\quad (\iota_m \psi)(t',x) = (U^{t',t}\, \psi)(x)\:, \]
which allows us to canonically identify each Hilbert space~$(\H_t, (.|.)_t)$ with~$(\H_m, (.|.))$.
We denote the restriction of a smooth Dirac wave function~$\psi \in C^\infty(\scrM, S\scrM)$ to the
hypersurface~$\scrN_t$ by~$\psi_{|t}$.

\begin{Lemma} \label{lemma44} For every~$\phi \in C^\infty_0(\scrM, S\scrM)$,
\begin{align}
(s^\wedge_m \phi)(t,x) &= -i \int_{-\infty}^t \Big(U^{t, t'}\! \big( \beta
\nuslsh \phi_{|t'} \big) \Big)(x)\: dt' \label{swrep} \\
(k_m \phi)(t,x) &= \frac{1}{2 \pi} \int_{-\infty}^\infty \Big(U^{t, t'}\! \big( \beta
\nuslsh \phi_{|t'} \big) \Big)(x)\: dt' \:. \label{kmrep}
\end{align}
\end{Lemma}
\Proof The Dirac operator can be written as
\[ \Dir = \beta^{-1} \nuslsh \left( i \partial_t - H_t \right) \:, \]
where~$H_t$ is a purely spatial operator acting on~$\H_t$ (the ``Hamiltonian'').
We apply the Dirac operator to the right side of~\eqref{swrep}, which we denote by~$F(t,x)$.
As the integrand in~\eqref{swrep} is a solution of the Dirac equation, only the derivative of the
limit of integration needs to be taken into account,
\[ (\Dir - m) F(t,x) = \big(\beta^{-1} \nuslsh(t,x) \big)
\big(U^{t, t} ( \beta \nuslsh \phi_{|t} ) \big)(x) \:. \]
Using that~$U^{t,t}$ is the identity, we conclude that
\[ (\Dir - m) F(t,x) = \phi(t,x) \:. \]
Hence~$F(t,x)$ satisfies the defining equation of the Green's operator~\eqref{Greendef}.
Moreover, it is obvious that~$F(t,x)$ vanishes if~$t$ is in the past of the
support of~$\phi$. The unique solution of the Cauchy problem gives the result.

Repeating the above argument for the advanced Green's operator gives
\[ (s^\vee_m \phi)(t,x) = i \int_t^\infty \Big(U^{t, t'}\! \big( \beta
\nuslsh \phi_{|t'} \big) \Big)(x)\: dt' \label{sarep} \:. \]
We finally apply~\eqref{kmdef} to obtain~\eqref{kmrep}.
\QED

For what follows, it is useful to identify~$\H_m$ with the Hilbert space~$\H_{t_0}$
for some fixed time~$t_0$. The formulas of the previous lemma are then rewritten
by multiplying with the time evolution operator. For example,
\beq \label{kfolirep}
k_m \phi= \frac{1}{2 \pi} \int_{-\infty}^\infty U^{t_0, t} \big( \beta
\nuslsh \,\phi \big)_{|t} \: dt \::\: C^\infty_0(\scrM, S\scrM) \rightarrow \H_{t_0}\:.
\eeq

\begin{Lemma}\label{sgn_operator_foliation} 
Assume that~$(\scrM,g)$ is $m$-finite. Then the fermionic signature operator~$\Sig$ as defined
by~\eqref{Sdef} has the representation
\[ \Sig = \frac{1}{2 \pi} \int_{-\infty}^\infty U^{t_0,t} \,(\beta \nuslsh)_{|t}\, U^{t,t_0} \:dt \;:\;
\H_{t_0} \rightarrow \H_{t_0}\:. \]
\end{Lemma}
\Proof Rewriting the space-time integral in~\eqref{stip} with Fubini's theorem and using the identity~$\nuslsh^2=\1$,
we obtain
\begin{align*}
\bra \phi_m | \psi_m \ket
&= \int_{-\infty}^\infty \bigg( \int_{\scrN_t} \Sl \phi_m | \psi_m \Sr_{(t,x)} \:\beta(t,x)\:d\mu_{\scrN_t}(x) \bigg)\: dt\\
&= \frac{1}{2 \pi} \int_{-\infty}^\infty (\phi_m \,|\, (\beta \nuslsh)_{|t} \,\psi_m)_t\: dt \\
&= \frac{1}{2 \pi} \int_{-\infty}^\infty (\phi_{|t_0} \,|\, U^{t_0,t} \,(\beta \nuslsh)_{|t}\, U^{t,t_0}
\,\psi_{|t_0})_{t_0}\: dt \:.
\end{align*}
Comparing with~\eqref{Sdef} gives the result.
\QED

Iterating~\eqref{kmrep}, we can make the following formal calculation,
\begin{align*}
(k_m\, k_m \,\phi)_{|t_0} &= \frac{1}{4 \pi^2} \int_{-\infty}^\infty dt
\int_{-\infty}^\infty dt' \:U^{t_0, t} \big( \beta \nuslsh \big)_{|t}
U^{t, t'}\! \big( \beta \nuslsh \phi \big)_{|t'} \\
&= \frac{1}{4 \pi^2} \int_{-\infty}^\infty dt
\int_{-\infty}^\infty dt' \:U^{t_0, t} (\beta \nuslsh)_{|t}\,
U^{t, t_0} \;U^{t_0, t'} (\beta \nuslsh \phi )_{|t'} \:,
\end{align*}
where in the second line we used the group property of the time evolution operator.
Comparing with~\eqref{kfolirep}, we obtain the simple relation
\[  k_m\, k_m \overset{\text{formally}}{=} \Sig\, k_m \:. \]
This is precisely the relation~\eqref{prodsugg} in the special case~$f,g \equiv 1$.
Iteration gives similar formal expressions for polynomials of~$k_m$, from
which~\eqref{prodsugg} can be obtained formally by approximation.
Although the last arguments are only formal, they explain how the functional
calculus~\eqref{prodsugg} comes about. In order to give this functional calculus a mathematical meaning,
one needs to evaluate weakly as is made precise by~\eqref{symm} and~\eqref{square}.

\subsection{Representation as a Distribution} \label{sec35}
We now represent the fermionic projector by a two-point distribution on~$\scrM$.
\begin{Thm} There is a unique distribution~${\mathcal{P}} \in \D'(\scrM \times \scrM)$ such that 
for all~$\phi, \psi \in C^\infty_0(\scrM, S\scrM)$,
\[ \bra \phi | P \psi \ket = {\mathcal{P}} \big( \overline{\phi} \otimes \psi \big)
\:. \]
\end{Thm}
\Proof According to Proposition~\ref{prpdual} and Definition~\ref{ferm_proj_Def},
\[ \bra \phi | P \psi \ket = (k_m \phi \,|\, P \psi) = - (k_m \phi \,|\, \chi_{(-\infty, 0)}(\Sig) \,k_m \psi) \:. \]
Since the norm of the operator~$\chi_{(-\infty, 0)}(\Sig)$ is bounded by one, we conclude that
\[ |\bra \phi | P \psi \ket| \leq \| k_m \phi\| \:\|k_m \psi\| 
= ( \bra \phi | k_m \phi \ket \: \bra \psi | k_m \psi \ket )^\frac{1}{2} \:, \]
where in the last step we again applied Proposition~\ref{prpdual}.
As~$k_m \in \D'(\scrM \times \scrM)$, the right side is continuous on~$\D(\scrM \times \scrM)$.
We conclude that also~$\bra \phi | P \psi \ket$ is continuous on~$\D(\scrM \times \scrM)$.
The result now follows from the Schwartz kernel theorem (see~\cite[Theorem~5.2.1]{hormanderI},
keeping in mind that this theorem applies just as well to bundle-valued distributions on a manifold
simply by working with the components in local coordinates and a local trivialization).
\QED

In order to get the connection to~\cite{PFP}, it is
convenient to use the standard notation with an integral kernel~$P(x,y)$,
\begin{align*}
\bra \phi | P \psi \ket &= \iint_{\scrM \times \scrM} \Sl \phi(x) \,|\, P(x,y) \,\psi(y) \Sr_x \: d\mu_\scrM(x)\: d\mu_\scrM(y) \\
(P \psi)(x) &= \int_{\scrM} P(x,y) \,\psi(y) \: d\mu_\scrM(y)
\end{align*}
(where~$P(.,.)$ coincides with the distribution~${\mathcal{P}}$ above).
In view of Proposition~\ref{prpPpm}, we know that last integral is not only a distribution,
but a function which is square integrable over every Cauchy surface.
Moreover, the symmetry of~$P$, \eqref{symmetry}, implies that
\[ P(x,y)^* = P(y,x) \:, \]
where the star denotes the adjoint with respect to the spin scalar product.
Finally, the spatial normalization property of Proposition~\ref{prpspatnorm} makes it possible
to obtain the following representation of the fermionic projector.
\begin{Prp} Let~$(\psi_j)_{j \in \N}$ be an orthonormal basis of the subspace~$\chi_{(-\infty, 0)}(\Sig)$
of the Hilbert space~$\H_m$. Then
\beq \label{Psum}
P(x,y) = -\sum_{j=1}^\infty |\psi_j(x) \Sr \Sl \psi_j(y)|
\eeq
with convergence in~$\D'(\scrM \times \scrM)$.
\end{Prp}
\Proof Being a projector on~$\chi_{(-\infty, 0)}(\Sig)$, the operator~$\Pi$ defined by~\eqref{Pidef}
has the representation~$\Pi = \sum_j |\psi_j)(\psi_j|$ and thus, in view of~\eqref{print},
\[ (\Pi \phi_m)(x) =2 \pi \sum_{j \in \N} \psi_j(x) \int_\scrN \Sl \psi_j(y) | \nuslsh \phi_m(y) \Sr_y\: d\mu_\scrN(y)\:. \]
Comparing with~\eqref{Pidef} and using that~$\phi_m$ can be chosen arbitrarily on~$\scrN$,
one sees that~\eqref{Psum} holds for all~$y \in \scrN$. Since the Cauchy surface~$\scrN$ can be
chosen to intersect any given space-time point, the result follows.
\QED

\section{Connection to the Framework of Causal Fermion Systems} \label{sec4}
We now explain the relation to the framework of causal fermion systems as
introduced in~\cite{rrev} (see also~\cite{lqg}).
In order to get into this framework, we need to introduce an ultraviolet regularization.
This is done most conveniently with so-called regularization operators.
\begin{Def}  A family~$({\mathfrak{R}}_\varepsilon)_{\varepsilon>0}$
of bounded linear operators on~$\H_m$ are called
{\bf{regularization operators}} if they have the following properties:
\begin{itemize}
\item[(i)] Solutions of the Dirac equation are mapped to continuous solutions,
\[ {\mathfrak{R}}_\varepsilon \::\: \H_m \rightarrow C^0(\scrM, S\scrM) \cap \H_m \]
\item[(ii)] For every~$\varepsilon>0$ and~$x \in \scrM$, there is a constant~$c>0$ such that
\beq \label{Rbound}
\|({\mathfrak{R}}_\varepsilon \psi_m)(x)\| \leq c\, \|\psi_m\| \qquad \forall\; \psi_m \in \H_m\:.
\eeq
(where the norm on the left is any norm on~$S_x\scrM$).
\item[(iii)] In the limit~$\varepsilon \searrow 0$, the regularization operators go over to the identity
with strong convergence of~${\mathfrak{R}}_\varepsilon$ and~${\mathfrak{R}}_\varepsilon^*$, i.e.
\beq \label{strong}
{\mathfrak{R}}_\varepsilon \psi_m, \;{\mathfrak{R}}_\varepsilon^* \psi_m
\xrightarrow{\varepsilon \searrow 0} \psi_m {\text{ in~$\H_m$}} \qquad \forall\; \psi_m \in \H_m\:.
\eeq
\end{itemize}
\end{Def} \noindent
There are many possibilities to choose regularization operators.
As a typical example, one can choose finite-dimensional
subspaces~$\H^{(\ell)} \subset \Cisc(\scrM, S\scrM) \cap \H_m$ which are an exhaustion of~$\H_m$ in the
sense that~$\H^{(0)} \subset \H^{(1)} \subset \cdots$ and~$\H_m = \overline{\cup_\ell \H^{(\ell)}}$.
Setting~$\ell(\varepsilon) = \max ( [0, 1/\varepsilon] \cap \N)$, we can introduce the
operators~${\mathfrak{R}}_\varepsilon$ as the orthogonal projection operators to~$\H^{(\ell(\varepsilon))}$.
An alternative method is to choose a Cauchy hypersurface~$\scrN$, to mollify the restriction~$\psi_m|_\scrN$
to the Cauchy surface on the length scale~$\varepsilon$, and to define~${\mathfrak{R}}_\varepsilon \psi_m$
as the solution of the Cauchy problem for the mollified initial data.

Given regularization operators~${\mathfrak{R}}_\varepsilon$, for any~$\varepsilon>0$
we introduce the {\em{particle space}} $(\H_\text{\tiny{particle}}, \la .|. \ra_{\H_\text{\tiny{particle}}})$ as
the Hilbert space
\[ \H_\text{\tiny{particle}} = \ker \big( \mathfrak{R}_\varepsilon\, \chi_{(-\infty, 0)}(\Sig) \big)^\perp\:,\qquad
\la .|. \ra_{\H_\text{\tiny{particle}}} = ( .|.)|_{\H_\text{\tiny{particle}} \times \H_\text{\tiny{particle}}} \:. \]
Next, for any~$x \in \scrM$ we consider the bilinear form
\[ b \::\: \H_\text{\tiny{particle}} \times \H_\text{\tiny{particle}} \rightarrow \C\:,\quad
b(\psi_m, \phi_m) = -\Sl ({\mathfrak{R}}_\varepsilon \,\psi_m)(x) \:|\:
({\mathfrak{R}}_\varepsilon \,\phi_m)(x) \Sr_x \:. \]
This bilinear form is bounded in view of~\eqref{Rbound}.
The {\em{local correlation operator}}~$F^\varepsilon(x)$ is defined as the signature operator
of this bilinear form, i.e.\
\[ b(\psi_m, \phi_m) = \la \psi_m \,|\, F^\varepsilon(x)\, \phi_m \ra_{\H_\text{\tiny{particle}}} \qquad
\text{for all~$\psi_m, \phi_m \in \H_\text{\tiny{particle}}$}\:. \]
Taking into account that the spin scalar product has signature~$(n,n)$,
the local correlation operator is a symmetric operator in~$\Lin(\H_\text{\tiny{particle}})$
of rank at most~$2n$,
which has at most $n$ positive and at most $n$ negative eigenvalues.
Finally, we introduce the {\em{universal measure}}~$d\rho= F^\varepsilon_* \,d\mu_\scrM$ as the push-forward
of the volume measure on~$\scrM$ under the mapping~$F^\varepsilon$
(thus~$\rho(\Omega) := \mu_\scrM((F^\varepsilon)^{-1}(\Omega))$).
Omitting the subscript ``particle'', we thus obtain
a causal fermion system as defined in~\cite[Section~1.2]{rrev}:

\begin{Def} 
Given a complex Hilbert space~$(\H, \la .|. \ra_\H)$ (the {\bf{``particle space''}})
and a parameter~$n \in \N$ (the {\bf{``spin dimension''}}), we let~$\F \subset \Lin(\H)$ be the set of all
self-adjoint operators on~$\H$ of finite rank, which (counting with multiplicities) have
at most~$n$ positive and at most~$n$ negative eigenvalues. On~$\F$ we are given
a positive measure~$\rho$ (defined on a $\sigma$-algebra of subsets of~$\F$), the so-called
{\bf{universal measure}}. We refer to~$(\H, \F, \rho)$ as a {\bf{causal fermion system}}.
\end{Def} \noindent
The formulation as a causal fermion system gives contact to a general mathematical framework
in which there are many inherent analytic and geometric structures
(see~\cite{continuum, lqg}).
In particular, the differential geometric objects of spin geometry have a 
canonical generalization to the regularized theory.
Namely, starting from a causal fermion system~$(\H, \F, \rho)$ one
defines space-time~$M$ as the support of the universal measure, $M := \supp \rho$.
Note that with this definition, the space-time points~$x, y \in M$ are operators on~$\H$
(thinking of our above construction of the causal fermion system,
this means that we identify a space-time point~$x$ with its local correlation
operator~$F^\varepsilon(x)$). 
On~$M$, we consider the topology induced by~$\F \subset \Lin(\H)$.
The causal structure is encoded in the spectrum
of the operator products~$x y$:
\begin{Def} For any~$x, y \in \F$, the product~$x y$ is an operator
of rank at most~$2n$. We denote its non-trivial eigenvalues
by $\lambda^{xy}_1, \ldots, \lambda^{xy}_{2n}$  (where we count with algebraic multiplicities).
The points~$x$ and~$y$ are called {\bf{timelike}} separated if the~$\lambda^{xy}_j$ are all real
and not all equal.
They are said to be {\bf{spacelike}} separated if all the~$\lambda^{xy}_j$ have the same absolute value.
In all other cases, the points~$x$ and~$y$ are said to be {\bf{lightlike}} separated.
\end{Def} \noindent
Next, we define the spin space~$S_x$ by~$S_x=x(\H) \subset \H$ endowed with the inner
product~$\Sl .|. \Sr_x := -\la .| x . \ra_\H$.
The kernel of the fermionic projector with regularization is introduced by
\beq \label{Pepsabs}
P^\varepsilon(x,y) = \pi_x \,y \::\: S_y \rightarrow S_x \:,
\eeq
where~$\pi_x$ is the orthogonal projection to~$S_x$ in~$\H$.
Connection and curvature can be defined as in~\cite[Section~3]{lqg}.
We remark for clarity that the Dirac equation and the bosonic field equations
(like the Maxwell or Einstein equations) cannot be formulated intrinsically
in a causal fermion system. Instead, as the main analytic structure one has
the causal action principle. We also point out that in the abstract framework,
it is impossible to perform the spatial integration in~\eqref{Pidef}. As a consequence,
it makes no sense to speak of the spatial normalization of the fermionic projector,
and the notion of a ``projector'' becomes unclear.
Therefore, in the abstract framework one refers to~\eqref{Pepsabs} as the kernel
of the fermionic {\em{operator}}. For a detailed discussion of the spatial normalization
in the context of causal fermion systems we refer to~\cite{norm}.

We conclude this section by deriving more explicit formulas for the local correlation
operators. Moreover, we compute the regularized fermionic projector and compare it
to the unregularized fermionic projector of Definition~\ref{ferm_proj_Def}.
To this end, for any~$x \in \scrM$ we define the {\em{evaluation map}}~$e_x^\varepsilon$ by
\beq \label{evalmap}
e^\varepsilon_x \::\: \H_m \rightarrow S_x\scrM \:,\qquad
e^\varepsilon_x \,\psi_m = ({\mathfrak{R}}_\varepsilon \,\chi_{(-\infty, 0)}(\Sig) \,\psi_m)(x)\:.
\eeq
We denote its adjoint by~$\iota^\varepsilon_x$,
\[ \iota^\varepsilon_x := (e^\varepsilon_x)^* \::\: S_x\scrM \rightarrow \H_m\:. \]
Multiplying~$\iota^\varepsilon_x$ by~$e^\varepsilon_x$ gives us
back the local correlation operator~$F^\varepsilon(x)$
(extended by zero to the orthogonal complement of~$\H_\text{\tiny{particle}}$),
\beq \label{Fepsdef}
F^\varepsilon(x) = - \iota^\varepsilon_x \,e^\varepsilon_x \::\: \H_m \rightarrow \H_m \:.
\eeq
Let us compute the adjoint of the evaluation map.
For any~$\xi \in S_x\scrM$ and~$\psi_m \in \H_m$, we have according to~\eqref{evalmap}
\[ ( (e^\varepsilon_x)^* \xi \,|\, \psi_m) 
= \Sl \xi \,|\, {\mathfrak{R}}_\varepsilon \,\chi_{(-\infty, 0)}(\Sig)\, \psi_m \Sr_x \\
= \bra \delta_x \xi \,|\, {\mathfrak{R}}_\varepsilon \,\chi_{(-\infty, 0)}(\Sig)\, \psi_m \ket \:, \]
where~$\delta_x$ is the $\delta$-distribution supported at~$x$ (thus in local coordinates,
$\delta_x(y) = |\det g(x)|^{-\frac{1}{2}}\, \delta^4(x-y)$). Applying Proposition~\ref{prpdual} gives
\[ ( (e^\varepsilon_x)^* \xi \,|\, \psi_m) = ( k_m \,\delta_x\, \xi \,|\, {\mathfrak{R}}_\varepsilon \,\chi_{(-\infty, 0)}(\Sig)\, \psi_m ) = ( \chi_{(-\infty, 0)}(\Sig)\, {\mathfrak{R}}_\varepsilon^*\, k_m \,\delta_x\, \xi \,|\, \psi_m ) \]
and thus
\beq \label{iotaform}
\iota^\varepsilon_x = (e^\varepsilon_x)^* = \chi_{(-\infty, 0)}(\Sig)\, {\mathfrak{R}}_\varepsilon^*\, k_m \,\delta_x\:.
\eeq
Combining this relation with~\eqref{evalmap} and~\eqref{Fepsdef}, the local correlation operator
takes the more explicit form
\[ F^\varepsilon(x) = -\iota^\varepsilon_x \,e^\varepsilon_x
= - \chi_{(-\infty, 0)}(\Sig)\, {\mathfrak{R}}_\varepsilon^*\, k_m \,\delta_x\,
{\mathfrak{R}}_\varepsilon \,\chi_{(-\infty, 0)}(\Sig) \:. \]

We next introduce the kernel of the regularized fermionic projector by
\beq \label{Pepsdef}
P^\varepsilon(x,y) = -e^\varepsilon_x \,\iota^\varepsilon_y \:.
\eeq
After suitably identifying the spinor spaces~$S_x\scrM$ and~$S_y\scrM$ with the
corresponding spin spaces~$S_x$ and~$S_y$, this definition indeed agrees with
the abstract definition~\eqref{Pepsabs} (for details see~\cite[Section~4.1]{lqg}).
Even without going through the details of this identification,
the definition~\eqref{Pepsdef} can be understood immediately by computing
the eigenvalues of the closed chain.
Starting from the definition~\eqref{Pepsabs}, the corresponding closed chain is
given by~$A^\varepsilon_{xy} := P^\varepsilon(x,y)\, P^\varepsilon(y,x) = \pi_x \,y \,x\, \pi_y$.
Keeping in mind that in~\eqref{Pepsabs} the space-time points are identified with the
corresponding local correlation matrices, this means that the spectrum of the closed chain 
is the same as that of the product~$ F(y)\, F(x)$ (except possibly for irrelevant zeros in the spectrum).
Taking the alternative definition~\eqref{Pepsdef} as the starting point,
the closed chain is given by
\[ A^\varepsilon(x,y) = (e^\varepsilon_x \,\iota^\varepsilon_y) \:(e^\varepsilon_y \,\iota^\varepsilon_x) \:. \]
Since a cyclic commutation of the operators has no influence on the eigenvalues, we conclude that the closed chain
is isospectral to the operator
\[ \iota^\varepsilon_y e^\varepsilon_y\, \iota^\varepsilon_x e^\varepsilon_x = F(y) \, F(x) \:, \]
giving agreement with the abstract definition~\eqref{Pepsabs}.

The corresponding regularized fermionic projector is defined by
\[ (P^\varepsilon(\phi))(x) = \int_\scrM P^\varepsilon(x,y)\: \phi(y)\: d\mu_\scrM(y) \:. \]
Using~\eqref{Pepsdef} together with~\eqref{iotaform} and~\eqref{evalmap},
this operator can be written as
\beq \label{Pepsform}
P^\varepsilon = - {\mathfrak{R}}_\varepsilon \,\chi_{(-\infty, 0)}(\Sig)\, {\mathfrak{R}}_\varepsilon^*\, k_m
\::\: C^\infty_0(\scrM, S\scrM) \rightarrow C^0(\scrM, S\scrM) \cap \H_m \:.
\eeq

The next proposition shows that if the regularization is removed,
the operator~$P^\varepsilon$ converges weakly to~$P$.
\begin{Prp} For every~$\phi, \psi \in C^\infty_0(\scrM, S\scrM)$,
\[ \bra \phi | P^\varepsilon \psi \ket \xrightarrow{\varepsilon \searrow 0} \bra \phi | P \psi \ket \:. \]
\end{Prp}
\Proof Applying Proposition~\ref{prpdual} and~\eqref{Pepsform}, we get
\[ \bra \phi | P^\varepsilon \psi \ket = -( k_m \phi | {\mathfrak{R}}_\varepsilon \,\chi_{(-\infty, 0)}(\Sig)\, {\mathfrak{R}}_\varepsilon^*\, k_m \psi)  =  -( {\mathfrak{R}}_\varepsilon^* \,k_m \phi | \chi_{(-\infty, 0)}(\Sig)\, {\mathfrak{R}}_\varepsilon^*\, k_m \psi) . \]
Now use that the operators~$R_\varepsilon^*$ converge strongly
according to~\eqref{strong}.
\QED

\section{Example: A Closed Friedmann-Robertson-Walker Universe} \label{secFRW}
We now want to complement the abstract construction of the fermionic projector by
a detailed analysis in a closed Friedmann-Robertson-Walker space-time. 
In so-called conformal coordinates, the line element reads
\begin{equation} \label{lineelement_FRW}
ds^2 = R(\tau)^2 \,\Big( d\tau^2 - d\chi^2 - \sin(\chi)^2\:(d\vartheta^2 + \sin^2{\vartheta} \:d\varphi^2) \Big) .
\end{equation}
Here $\tau \in (0,\pi)$ is a time coordinate, $\varphi \in [0, 2 \pi)$ and $\vartheta \in (0, \pi)$ are angular coordinates, and~$\chi\in (0,\pi)$ is a radial coordinate. The scale function~$R(\tau)$
should have the following properties.
We assume that~$\tau=0$ and~$\tau=\pi$ are the big bang and big crunch singularities, respectively.
This implies that
\[ R(0) = 0 = R(\pi) \qquad \text{and} \qquad R|_{(0, \pi)} > 0 \:. \]
Moreover, we assume that~$R$ is a $C^2$-function which is {\em{piecewise monotone}} 
(i.e.,\ the interval~$(0, \pi)$ can be divided into a finite number of subintervals on which~$R$ is monotone).
It is convenient to write the scale function as
\beq \label{scale function}
R(\tau)=R_{\max}\, g(\tau) \qquad \text{with} \qquad R_{\max} := \max_{(0, \pi)} R \:.
\eeq
A special case is the dust matter model~$R(\tau)= R_{\max} \,(1-\cos(\tau))$ (see~\cite[Section~5.3]{hawking+ellis}).

The spatial dependence of the Dirac equation can be separated
by eigenfunctions of the Dirac operator on~$S^3$ corresponding to the
eigenvalues~$\lambda \in \{\pm \frac{3}{2}, \pm \frac{5}{2}, \ldots \}$
(for details see~\cite{moritz}). After this separation, the time evolution
operator~$U^{\tau,\tau_0} \, \in \, C^1((0,\pi), \U(\C^2))$ of the Dirac equation 
is given as the solution of the initial value problem
\begin{align} \label{evolution_eqn_FRW}
i \partial_\tau U^{\tau,\tau_0} &= \left[ m R(\tau)
\begin{pmatrix} 1 & 0 \\ 0 & -1 \end{pmatrix} - \lambda \begin{pmatrix} 0 & 1 \\ 1 & 0 \end{pmatrix} \right]U^{\tau,\tau_0} \\
U^{\tau_0,\tau_0} &=\1_2. \label{initval}
\end{align}
According to Definition~\ref{ferm_proj_Def} and~\eqref{Ppmdef} as well as~\eqref{kfolirep}, we have
\begin{align}
P &= -\chi_{(-\infty, 0)}(\Sig)\, k_m \label{PFRW} \\
k_m(\phi) &= \frac1{2\pi} \int_0^\pi (U^{\tau, \tau_0})^* 
\begin{pmatrix}1 & 0 \\ 0 & -1 \end{pmatrix} \phi(\tau) \:R(\tau)\: d\tau\:, \label{kmFRW}
\end{align}
where~$\phi \in C^\infty_0((0, \pi), \C^2)$.

In the subsequent estimates, we shall work with the WKB approximation introduced as follows
(for more details see~\cite{moritz}). We first define~$V(\tau)$ as
a unitary matrix which diagonalizes the coefficient matrix in~\eqref{evolution_eqn_FRW}, i.e.\
\beq \label{Vdef}
V \begin{pmatrix} Rm & -\lambda \cr -\lambda & -Rm \end{pmatrix} V^{-1} = f 
\begin{pmatrix} 1 & 0 \\ 0 & -1 \end{pmatrix} ,
\eeq
where
\beq \label{fdef}
f(\tau) := \sqrt{\lambda^2+m^2R(\tau)^2} \:.
\eeq
We now introduce the WKB approximation by
\beq \label{Udef}
U^{\tau, \tau_0}_{\text{\tiny{WKB}}} = V(\tau)^{-1} \left(\begin{array}{cc} 
\displaystyle \exp\left(-i\int_{\tau_0}^\tau f \right) & 0 \cr 0 &
\displaystyle \exp\left(i\int_{\tau_0}^\tau f \right)   \end{array}\right) V(\tau_0) \:.
\eeq
Note that for all $\tau, \tau_0 \in(0,\pi)$, the matrices $U^{\tau, \tau_0},\,V(\tau)$ and $U^{\tau, \tau_0}_\text{\tiny{WKB}}$ are unitary.

Applying Lemma \ref{sgn_operator_foliation}, 
the signature operator~$\Sig$ as defined by~\eqref{Sdef} takes the form
\beq
\Sig = \int_0^\pi U^{\tau_0, \tau}_m \begin{pmatrix} 1 & 0 \\ 0 & -1 \end{pmatrix} U^{\tau, \tau_0}_m
\:R(\tau) \,d\tau \label{S_FRW} \:.
\eeq
Replacing the time evolution by the WKB approximation, we obtain the signature operator
\beq
\Sig_\text{\tiny{WKB}} = \int_0^\pi U^{\tau_0, \tau}_\text{\tiny{WKB}} \begin{pmatrix} 1 & 0 \\ 0 & -1 \end{pmatrix} U^{\tau, \tau_0}_\text{\tiny{WKB}} \,R(\tau)\, d\tau \:. \label{S_WKB}
\eeq
In analogy to~\eqref{PFRW} and~\eqref{kmFRW}, we introduce the fermionic projector
in the WKB approximation by
\begin{align}
P_\text{\tiny{WKB}} &= -\chi_{(-\infty, 0)}(\Sig_\text{\tiny{WKB}}) \, k_\text{\tiny{WKB}} \label{ferm_proj_FRW} \\
k_{\text{\tiny{WKB}}}(\phi) &= \frac1{2\pi} \int_0^\pi (U^{\tau, \tau_0}_\text{\tiny{WKB}})^* 
\begin{pmatrix}1 & 0 \\ 0 & -1 \end{pmatrix} \phi(\tau) \:R(\tau)\, d\tau \label{kmWKB}\:.
\end{align}

In the following two theorems, we specify under which conditions and in which sense
the fermionic projector is well-approximated by WKB wave functions.
We first state the theorems and discuss them afterwards.
\begin{Thm} \label{thmPWKB} For given~$\tau_0 \in (0, \pi)$ and a given function~$g$, the
function~$P_\text{\tiny{\rm{WKB}}}$ as defined by~\eqref{ferm_proj_FRW}
can be represented for any values of the parameters~$\lambda$, $m$ and~$R_{\text{max}}$ by
\begin{align*}
P_\text{\tiny{\rm{WKB}}}(\phi) &= -\frac{1}{2 \pi}
\int_0^\pi V(\tau_0)^{-1} \begin{pmatrix} 
0 & 0 \cr 0 &
\displaystyle \exp\left(i\int_\tau^{\tau_0} f \right) \end{pmatrix}
V(\tau) \begin{pmatrix}1 & 0 \\ 0 & -1 \end{pmatrix} \phi(\tau) \:R(\tau)\, d\tau \\
&\qquad \times \bigg(1 + \O \bigg( \frac{\sqrt{\lambda^2 + m^2 R_{\max}^2}}{m^2 R_{\max}^2} \bigg) \bigg) \,.
\end{align*}
\end{Thm}

\begin{Thm} \label{thmPmP}
For any constant~$K>0$, there is a constant~$c$ (only depending on~$K$, $\tau_0$ and the function~$g$),
such that for all~$m$ and~$R_{\max}$ with~$m R_{\max}>1$ the following statement holds:
For every~$\lambda$ in the range
\beq \label{lames}
|\lambda| \leq K\, m R_{\max}
\eeq
and every~$\phi \in C^\infty_0((0, \pi), \C^2)$, we have the estimate
\beq \label{Perr}
\|(P - P_\text{\tiny{\rm{WKB}}})(\phi)\| \leq c\, (m R_{\max})^{-\frac{1}{5}}\; R_{\max} \int_0^\pi \|\phi(\tau)\|\,
d\tau \:.
\eeq
\end{Thm} \noindent
Comparing the exponential factors in~\eqref{Udef} with those in Theorem~\ref{thmPWKB},
one sees that $P_\text{\tiny{WKB}}$ only involves the factor~$\exp(i \int f)$, whereas
the factor~$\exp(-i \int f)$ in~\eqref{Udef} has disappeared. In this sense, our formula
of~$P_\text{\tiny{WKB}}$ only involves the {\em{negative frequency}} solutions of the Dirac equation.
Thus this formula corresponds precisely to the naive picture of the Dirac sea as being composed
of all negative-energy solutions of the Dirac equation.
Theorem~\ref{thmPWKB} and Theorem~\ref{thmPmP} show that the fermionic projector
agrees with this naive picture, up to error terms which we now discuss.
We first point out that, according to~\eqref{PFRW} and~\eqref{kmFRW}, the fermionic
projector has the naive scaling
\[ P(\phi) \sim \int_0^\pi \phi(\tau)\: R(\tau)\, d\tau \:. \]
In order to compare with the error estimate~\eqref{Perr}, we need to assume that~$\phi$
is supported away from the big bang and big crunch singularities, so that
\beq \label{singaway}
\int_0^\pi \phi(\tau)\: R(\tau)\, d\tau \sim R_{\max} \int_0^\pi \phi(\tau)\, d\tau \:.
\eeq
This assumption is reasonable because we cannot expect the WKB approximation
to hold near the singularities (in particular because ``quantum oscillations'' become
relevant; see~\cite{bounce}).
Under this assumption, the estimate~\eqref{Perr} can be translated to a relative error of the
order~$\O((m R_{\max})^{-\frac{1}{5}})$.
We conclude that the error terms are under control provided that the size of the universe
is much larger than the Compton scale~$1/m$.
One should keep in mind that our theorems hold for a fixed function~$g$ in~\eqref{scale function}.
This implies that the metric must be nearly constant on the Compton scale.
Note that our estimates do not involve time integrals over the error, as one would get in a Gr\"onwall
estimate. This means that the local errors in different regions of space-time do not add up;
we merely need to keep the error small at every space-time point.
We also point out that, even when evaluating away from the singularities (see~\eqref{singaway}),
the behavior of the metric near the singularities still enters our construction
via the integral~\eqref{S_FRW}. It is a main point of our analysis to estimate this integral
without making any assumptions on the asymptotic form of~$g$
near the big bang or big crunch singularities.

We finally discuss how our estimates depend on the momentum~$\lambda$.
In view of~\eqref{lames} and the error term in Theorem~\ref{thmPWKB},
we may choose the quotient~$|\lambda|/(mR_{\max})$ arbitrarily large.
This makes it possible to even describe ultrarelativistic Dirac particles.
However, the constant~$c$ in~\eqref{Perr} and the error term in Theorem~\ref{thmPWKB}
depend on this quotient. This means that we
cannot take the limit~$|\lambda| \rightarrow \infty$ for fixed~$mR_{\max}$.
It is not clear whether in this limit, the WKB approximation of~$P$ really breaks down
or whether our estimates are simply not good enough to give a proper description
of the corresponding asymptotic behavior.

\subsection{Computation of~$\Sig_\text{\tiny{WKB}}$ and~$P_\text{\tiny{WKB}}$}
We now derive asymptotic formulas for $\Sig_\text{\tiny{WKB}}$ and~$P_\text{\tiny{WKB}}$
including error estimates.
\begin{Prp} For any~$\tau_0 \in (0, \pi)$ there is a constant~$c$ which depends only on~$\tau_0$
and the function~$g$ such that the matrix~$\Sig_\text{\tiny{\rm{WKB}}}$ as defined by~\eqref{S_WKB}
has the explicit approximation
\beq \label{SWKBrep}
\Sig_\text{\tiny{\rm{WKB}}} = \bigg( \int_0^\pi \frac{m R(\tau)^2}{f(\tau)}\: d\tau \bigg)\:
V(\tau_0)^{-1} \begin{pmatrix} 1 & 0 \\ 0 & -1 \end{pmatrix} V(\tau_0) + E
\eeq
with an error term~$E$ bounded by
\beq \label{Err}
\|E\| \leq \frac{c}{m}
\eeq
(here~$\| . \|$ is some norm on $2 \times 2$-matrices).
Moreover, the eigenvalues~$\mu^\pm_\text{\tiny{WKB}}$ of the matrix~$\Sig_\text{\tiny{WKB}}$ are given by
\beq\label{eigval_S_WKB}
\mu^\pm_\text{\tiny{WKB}} = \pm \sqrt{
\left( \lambda \int_0^\pi \frac{\cos\phi}{f}\:R \,d\tau \right)^2 \!\!\!+
\left( \lambda \int_0^\pi \frac{\sin\phi}{f}\:R \,d\tau \right)^2 \!\!\! +\left( m \int_0^\pi \frac{R^2}{f}\:d\tau \right)^2} \:,
\eeq
where
\beq \label{phidef}
\phi(\tau) := -2\int_{\tau_0}^\tau \sqrt{\lambda^2 + m^2R^2} \:.
\eeq
\end{Prp}
\Proof
A straightforward computation gives
\begin{align*}
(U^{\tau, \tau_0}_\text{\tiny{WKB}})^* & \begin{pmatrix} 1 & 0 \\ 0 & -1 \end{pmatrix} U^{\tau,
\tau_0}_\text{\tiny{WKB}} \,R(\tau)
= \frac{m R(\tau)^2}{f(\tau)}\: \frac{1}{f(\tau_0)} \begin{pmatrix} m R(\tau_0) & -\lambda
\\ -\lambda & -m R(\tau_0) \end{pmatrix} \\
&+ \frac{R(\tau)}{f(\tau)}\: \cos \phi \: \frac{\lambda}{f(\tau_0)} \begin{pmatrix} \lambda & m R(\tau_0)
\\ m R(\tau_0) & -\lambda \end{pmatrix} 
+ \frac{R(\tau)}{f(\tau)}\: \sin \phi\:\lambda \begin{pmatrix} 0 & -i \\ i & 0 \end{pmatrix} .
\end{align*}
Carrying out the integral in~\eqref{S_WKB}, we can compute the eigenvalues of the
resulting matrix to obtain~\eqref{eigval_S_WKB}.
In order to derive asymptotic formulas, one must keep in mind that the
factors~$\sin \phi$ and~$\cos \phi$ oscillate,
resulting in small contributions to~$\Sig_{\text{\tiny{WKB}}}$. 
Let us quantify this effect for the integral involving~$\cos \phi$ (for the integral involving~$\sin \phi$
the argument is exactly the same). We first transform the integral by
\[ \int_0^\pi \frac{R(\tau)}{f(\tau)}\: \cos \phi \: d\tau
= -\int_0^\pi \frac{R(\tau)}{f(\tau) \, \phi'(\tau)}\: \frac{d}{d\tau} \sin \phi \: d\tau
= \int_0^\pi \frac{R(\tau)}{2 f(\tau)^2}\: \frac{d}{d\tau} \sin \phi \: d\tau \:. \]
Integrating by parts and using that~$R$ vanishes at both end points, we obtain
\[ \int_0^\pi \frac{R(\tau)}{f(\tau)}\: \cos \phi \: d\tau
= - \int_0^\pi \frac{ \dot{R}\:(\lambda^2-m^2 R^2) } {(\lambda^2+m^2R^2)^2} \,\sin\phi\, d\tau \:. \]
This yields the estimate
\[ \left| \int_0^\pi \frac{R(\tau)}{f(\tau)}\: \cos \phi \: d\tau \right|
\leq \int_0^\pi \left| \frac{ \dot{R}} {\lambda^2+m^2R^2} \right| d\tau 
= \frac{1}{|\lambda m|} \int_0^\pi \left| \frac{d}{d\tau} \arctan \left( \frac{m R}{\lambda} 
\right) \right| d\tau \:. \]
On an interval where~$R$ is monotone, we can carry out the last integral,
giving at most~$\pi/2$. Since~$R$ is piecewise monotone, we can subdivide the interval~$(0, \pi)$
into~$N$ subintervals on which~$R$ is monotone and carry out the integral on each such subinterval.
We conclude that
\[ \left| \int_0^\pi \frac{R(\tau)}{f(\tau)}\: \cos \phi \: d\tau \right| \leq \frac{1}{|\lambda m|}\: \frac{N \pi}{2}\:. \]
Next, a direct calculation shows that the matrix
\[ \frac{1}{f(\tau_0)} \begin{pmatrix} \lambda & m R(\tau_0)
\\ m R(\tau_0) & -\lambda \end{pmatrix} \]
has eigenvalues~$\pm 1$ and is thus uniformly bounded. This completes the proof.
\QED

\Proof[Proof of Theorem~\ref{thmPWKB}]
Writing the spectral calculus with residues, we have
\[ -\chi_{(-\infty, 0)}(\Sig_\text{\tiny{WKB}})
= \frac{1}{2 \pi i} \ointctrclockwise_\Gamma (\Sig_{\text{\tiny{WKB}}}-\lambda)^{-1}\: d\lambda \:, \]
where~$\Gamma$ is a contour which encloses the negative eigenvalue of~$\Sig_\text{\tiny{WKB}}$.
Estimating the integral in~\eqref{SWKBrep} by
\[ \int_0^\pi \frac{m R^2}{\sqrt{\lambda^2 + m^2 R^2}}\: d\tau
\geq \int_0^\pi \frac{m R^2}{\sqrt{\lambda^2 + m^2 R_{\max}^2}}\: d\tau
= c \: \frac{m R_{\max}^2}{\sqrt{\lambda^2 + m^2 R_{\max}^2}} \]
with
\[ c:= \int_0^\pi g^2 \, d\tau > 0 \:, \]
we find that~$\Gamma$ can be chosen as a circle with center~$\mu^-_{\text{\tiny{WKB}}}$
and radius~$r$ given by
\[ r = c \: \frac{m R_{\max}^2}{\sqrt{\lambda^2 + m^2 R_{\max}^2}} \:. \]
Denoting the first summand in~\eqref{SWKBrep} by~$\Sig^{(0)}_{\text{\tiny{WKB}}}$
and computing the contour integral gives
\[  -\chi_{(-\infty, 0)} \big( \Sig^{(0)}_\text{\tiny{WKB}} \big)
= V(\tau_0)^{-1} \begin{pmatrix} 0 & 0 \\ 0 & -1 \end{pmatrix} V(\tau_0) \:. \]
In order to estimate the error term~$E$ in~\eqref{SWKBrep}, we write the corresponding contour integrals as
\begin{align*}
\ointctrclockwise_\Gamma &
\left[ (\Sig_\text{\tiny{WKB}} - \lambda)^{-1} - (\Sig_\text{\tiny{WKB}}^{(0)} - \lambda)^{-1} \right] d\lambda
= \ointctrclockwise_\Gamma \int_0^1 \frac{d}{dt}  (\Sig_\text{\tiny{WKB}}^{(0)} + t E - \lambda)^{-1} \:dt\, d\lambda \\
&= -\ointctrclockwise_\Gamma \int_0^1 (\Sig_\text{\tiny{WKB}}^{(0)} + t E - \lambda)^{-1}
\,E\, (\Sig_\text{\tiny{WKB}}^{(0)} + t E - \lambda)^{-1}  \:dt\, d\lambda \:.
\end{align*}
Taking the absolute value and estimating the integrand, we obtain the error bound
\[ \frac{c}{2 \pi} \ointctrclockwise_\Gamma \frac{\|E\|}{r^2} \:d|\lambda| = \frac{c\, \|E\|}{r}
\leq c \:\frac{\sqrt{\lambda^2 + m^2 R_{\max}^2}}{m^2 R_{\max}^2} \:, \]
where in the last step we applied~\eqref{Err}.
Using~\eqref{ferm_proj_FRW}, \eqref{kmWKB} and~\eqref{Udef} gives the result.
\QED

\subsection{Estimates of~$U-U_\text{\tiny{WKB}}$ and~$\Sig-\Sig_\text{\tiny{WKB}}$}
The goal of this section is to derive the following estimate.
\begin{Prp} \label{bound_|S-SWKB|_Lemma}\label{prp1}
For any~$\tau_0 \in (0, \pi)$ there is a constant~$c$ which depends only on~$\tau_0$ and the function~$g$
such that for all~$m$ and~$R_{\max}$ with~$m R_{\max}>1$,
\beq \label{bound_|S-SWKB|_Lemma_eqn}
\lVert \Sig-\Sig_\text{\tiny{\rm{WKB}}} \rVert < c \: m^{-\frac{1}{5}} R_{\max}^{\frac{4}{5}}
\eeq
(where~$\|.\|$ again denotes a matrix norm).
\end{Prp}

In preparation, we begin with three technical lemmas. Note that, as the
matrices~$U^{\tau, \tau_0}$ and~$U^{\tau, \tau_0}_\text{\tiny{WKB}}$ are both unitary,
instead of~$U^{\tau, \tau_0}-U^{\tau, \tau_0}_\text{\tiny{WKB}}$
we can just as well estimate the matrix~$W(\tau) - \1$, where~$W$ is the 
unitary matrix
\beq \label{Wdef}
W(\tau) := (U^{\tau, \tau_0}_\text{\tiny{WKB}})^* \,U^{\tau, \tau_0} \:.
\eeq
A short calculation using~\eqref{evolution_eqn_FRW}, \eqref{Vdef} and~\eqref{Udef}
shows that
\[ \partial_\tau W(\tau) = (U^{\tau, \tau_0}_\text{\tiny{WKB}})^* \,V(\tau)^* \:(\partial_\tau V(\tau))\: U^{\tau, \tau_0}
\:.\]
Again using the definition of~$W(\tau)$, we obtain the differential equation
\beq \label{Weq}
\partial_\tau W(\tau) = X(\tau)\, W(\tau) 
\qquad \text{with} \qquad X := (U^{\tau, \tau_0}_\text{\tiny{WKB}})^* \,V^* (\partial_\tau V)\, U^{\tau, \tau_0}_\text{\tiny{WKB}}\:.
\eeq
A straightforward computation gives
\beq \label{Xdef}
X = \frac{\lambda m \dot{R}}{2 f^2} \: \frac{1}{f_0}
\begin{pmatrix} -i \lambda \sin(\phi) & f_0 \cos(\phi) - i m R_0 \sin(\phi) \\
-f_0 \cos(\phi) - i m R_0 \sin(\phi) & i \lambda \sin(\phi) \end{pmatrix} ,
\eeq
where~$\phi$ is again the function~\eqref{phidef} and
\[ f_0:=f(\tau_0) \:,\qquad R_0 := R(\tau_0)\:. \]

\begin{Lemma} \label{lemmaes1}
Assume that the function~$R(\tau)$ is monotone on the interval~$[\tau_1, \tau_2] \subset (0, \pi)$.
Then
\[ \left| \|W(\tau)-\1\| \Big|_{\tau_1}^{\tau_2} \right|
\leq \frac{1}{2} \left| \,\arctan \!\Big( \frac{m R(\tau)}{\lambda} \Big) \Big|_{\tau_1}^{\tau_2} \right| . \]
\end{Lemma}
\Proof Using Kato's inequality together with the fact that~$W$ is unitary, we know from~\eqref{Weq} that
\[ \| W - \1 \| \Big|_{\tau_1}^{\tau_2} \leq \int_{\tau_1}^{\tau_2} \| \partial_t W(\tau) \|\, d\tau
\leq \int_{\tau_1}^{\tau_2} \| X(\tau) \|\, d\tau\:. \]
The matrix appearing on the right hand side of~\eqref{Xdef} is anti-Hermitian with eigenvalues~$\pm i f_0$. Hence
\beq \label{Xes}
\| X \| \leq \left| \frac{\lambda m \dot{R}}{2 f^2} \right| 
= \left| \frac{1}{2} \frac{d}{d\tau} \arctan \!\Big( \frac{m R}{\lambda} \Big) \right| ,
\eeq
where the last step is immediately verified by computing the derivative of the $\arctan$
and using~\eqref{fdef}. Integrating on both sides and using that~$R$ is monotone gives the result.
\QED

\begin{Lemma} \label{lemma63} For any~$\tau_0 \in (0, \pi)$ there is a constant~$c$ depending only on~$\tau_0$
and the function~$g$ such that
\begin{align}
\|W(\tau) - \1 \| &\leq \frac{c \,|\lambda|}{m R(\tau)} \label{Wes} \\
\int_0^\pi \|W(\tau)-\1\|\, R(\tau)\: d\tau &\leq \frac{c \pi \,|\lambda|}{m}\:. \label{intW}
\end{align}
\end{Lemma}
\Proof The inequality~\eqref{intW} follows immediately by integrating~\eqref{Wes}.
For the proof of~\eqref{Wes}, it suffices to consider the case~$\tau>\tau_0$, because the
case~$\tau < \tau_0$ is analogous.
We choose intermediate points~$\tau_1, \ldots, \tau_N$ with
\[ \tau_0 < \tau_1 < \cdots < \tau_N = \pi \:, \]
such that~$R$ restricted to the subintervals~$[\tau_{\ell-1}, \tau_\ell]$ is monotone for all~$\ell=1, \ldots N$.
Then~$\tau$ lies in one of the subintervals, $\tau \in [\tau_{n-1}, \tau_n]$.
Applying Lemma~\ref{lemmaes1} on the interval~$[\tau_0, \tau_1]$ and using that~$W(\tau_0)=\1$,
we obtain
\begin{align*}
2\, \| W(\tau_1)-\1\|
& \leq \left| \,\arctan \!\Big( \frac{m R(\tau)}{\lambda} \Big) \Big|_{\tau_0}^{\tau_1} \right| \\
&\leq \left( \frac{\pi}{2} - \arctan \!\Big( \frac{m R(\tau_0)}{\lambda} \Big) \right)
+  \left( \frac{\pi}{2} - \arctan \!\Big( \frac{m R(\tau_1)}{\lambda} \Big) \right) .
\end{align*}
Applying the elementary inequality
\[ \frac{\pi}{2} - \arctan(x) \leq \frac{1}{x} \qquad \text{for all~$x>0$} \]
gives
\[ 2\, \| W(\tau_1)-\1\| \leq \frac{|\lambda|}{m R(\tau_0)} + \frac{|\lambda|}{m R(\tau_1)} \:. \]
Proceeding similarly on the other intervals, we conclude that
\[ \| W(\tau)-\1\| \leq \frac{|\lambda|}{m R(\tau_0)} + \cdots + \frac{|\lambda|}{m R(\tau_{n-1})}
\:+\: \frac{|\lambda|}{m R(\tau)} \:. \]
Using the scaling~\eqref{scale function}, we obtain
\[ \| W(\tau)-\1\| \leq \frac{|\lambda|}{m R(\tau)}\:\left( \frac{1}{g(\tau_0)} + \cdots + \frac{1}{g(\tau_N)} + 1 \right) , \]
giving the result.
\QED

\begin{Lemma}\label{lemmaes3} Suppose that~$\lambda > (m R_{\max})^{\frac{4}{5}}$. Then
there is a constant~$c$ which depends only on~$\tau_0$ and the function~$g$ such that
for all~$m$ and~$R_{\max}$ with~$m R_{\max}>1$,
\begin{align}
\| W(\tau)-\1 \| &\leq c\: (m R_{\max})^{-\frac{1}{5}} \label{esW} \\
\int_0^\pi \|W-\1\|\, R(\tau)\: d\tau &\leq c\, m^{-\frac{1}{5}} R_{\max}^{\frac{4}{5}}\:. \label{Wint}
\end{align}
\end{Lemma}
\Proof We write~\eqref{Xdef} as
\beq \label{Xrep}
X(\tau) = h(\tau)\: \frac{d}{d\tau} M(\tau) \:,
\eeq
where
\begin{align*}
h &= \frac{\lambda m \dot{R}}{4 f^3} \qquad \text{and} \\
M &=  \frac{1}{f_0} \begin{pmatrix} -i \lambda \cos(\phi) & -f_0 \sin(\phi) -i m R_0 \cos(\phi) \\
f_0 \sin(\phi) - i m R_0 \cos(\phi) & i \lambda \cos(\phi) \end{pmatrix} .
\end{align*}
Integrating~\eqref{Weq}, we can employ~\eqref{Xrep} and integrate by parts to obtain
\begin{align*}
(W(\tau)-\1) \Big|_{\tau_1}^{\tau_2} &= \int_{\tau_1}^{\tau_2} X W\: d\tau
= h M W \Big|_{\tau_1}^{\tau_2} -\int_{\tau_1}^{\tau_2} M\: \frac{d}{d\tau} (h W )\: d\tau \\
&= h M W \Big|_{\tau_1}^{\tau_2} - \int_{\tau_1}^{\tau_2} \left( M \dot{h} W 
+ M h X W \right) d\tau\:,
\end{align*}
where in the last line we used~\eqref{Weq}.
The matrix~$M$ is anti-Hermitian and has the eigenvalues~$\pm i$.
Moreover, the matrix~$X$ can be estimated by the first inequality in~\eqref{Xes},
which we now write as~$\|X\| \leq |2 f h|$. Using furthermore
that~$W$ is unitary, we obtain
\beq \label{PI}
\left| \| W(\tau)-\1 \| \Big|_{\tau_1}^{\tau_2} \right|
\leq \big| h(\tau_2) \big| + \big| h(\tau_1) \big|
 + \int_{\tau_1}^{\tau_2} \left( \big| \dot{h} \big| + \big| 2 f h^2 \big| \right) d\tau\:.
\eeq

Now suppose that~$|\lambda| \geq (m R_{\max})^\beta$ with~$\beta<1$ (choosing~$\beta=4/5$ later will
give the result). Using the estimate~$f \geq \lambda$, we obtain
\begin{align*}
|h| &\leq \frac{\lambda m}{\lambda^3}\: R_{\max} \, |\dot{g}|
\leq (m R_{\max})^{1-2 \beta}\: |\dot{g}| \\
\big| 2 f {h}^2 \big| &\leq \frac{m^2}{8 \lambda^3}\: R_{\max}^2 \:|\dot{g}|^2
\leq \frac{1}{8}\, (m R_{\max})^{2-3 \beta}\: |\dot{g}|^2 \\
\big| \dot{h} \big| &\leq \frac{1}{4}\, (m R_{\max})^{1-2 \beta}\: |\ddot{g}| + \frac{3}{4}\, (m R_{\max})^{3-4 \beta} \:|\dot{g}|^2\:.
\end{align*}
Using these inequalities in~\eqref{PI} for~$\tau_1 = \tau_0$ or~$\tau_2=\tau_0$,
we conclude that there is a constant~$c$ as in the statement of
the lemma such that
\[ \| W(\tau)-\1 \| \leq c\: (m R_{\max})^{3-4 \beta}\:. \]
Choosing~$\beta=4/5$ gives~\eqref{esW}. Integrating yields~\eqref{Wint}.
\QED

\Proof of Proposition \ref{bound_|S-SWKB|_Lemma}.
We first derive a bound on the norm of~$\Sig-\Sig_\text{\tiny{WKB}}$ in terms of $\|U-U_\text{\tiny{WKB}}\|$.
Introducing the notation $Y(\tau)=U^{\tau, \tau_0}
(U^{\tau, \tau_0}_\text{\tiny{WKB}}(\tau))^*$ and applying \eqref{S_FRW} and \eqref{S_WKB}, we find
\beq\label{bound_|S-SWKB|_eqn1}
\Sig-\Sig_\text{\tiny{WKB}} = \int^\pi_0 (U^{\tau, \tau_0})^* \left[ \begin{pmatrix} 1 & 0 \\ 0 & -1 \end{pmatrix} - Y(\tau)
\begin{pmatrix} 1 & 0 \\ 0 & -1 \end{pmatrix} Y(\tau)^* \right] U^{\tau, \tau_0}\:
R(\tau)\, d\tau.
\eeq
Since $U$ and $Y$ are unitary matrices, \eqref{bound_|S-SWKB|_eqn1} implies that
\begin{align*}
\big\| \Sig&-\Sig_\text{\tiny{WKB}} \big\| \leq
\int^\pi_0 \Big\| \begin{pmatrix} 1 & 0 \\ 0 & -1 \end{pmatrix} - Y(\tau) \begin{pmatrix} 1 & 0 \\ 0 & -1 \end{pmatrix} Y(\tau)^*  \Big\| \:R(\tau) \,d\tau \cr
&= \int^\pi_0 \Big\| \big(\1-Y(\tau) \big) \begin{pmatrix} 1 & 0 \\ 0 & -1 \end{pmatrix}
+ Y(\tau) \begin{pmatrix} 1 & 0 \\ 0 & -1 \end{pmatrix} \left( \1 - Y(\tau)^* \right)  \Big\| \:R(\tau)\, d\tau \cr
&\leq  \int^\pi_0 \left(  \big\| \1 - Y(\tau) \big\| + \big\| \1 - Y(\tau)^*  \big\| \right) R(\tau)\, d\tau \cr
&= 2\int^\pi_0  \big\| U^{\tau, \tau_0} - U^{\tau, \tau_0}_\text{\tiny{WKB}}(\tau) \big\| R(\tau) \:d\tau
= 2\int^\pi_0  \big\| W(\tau) - \1 \big\| \:R(\tau) \:d\tau \:.
\end{align*}
Now, Lemma \ref{lemma63} yields~\eqref{bound_|S-SWKB|_Lemma_eqn}, while Lemma \ref{lemmaes3} gives the remaining case. This completes the proof.
\QED

\subsection{An Estimate of~$P-P_\text{\tiny{WKB}}$}

\Proof[Proof of Theorem~\ref{thmPmP}]
Introducing the abbreviations
\[ \mathcal{N}=-\chi_{(-\infty, 0)}(\Sig)
\qquad \text{and} \qquad \mathcal{N}_\text{\tiny{WKB}}=  -\chi_{(-\infty, 0)}(\Sig_\text{\tiny{WKB}}) \:, \]
we obtain from~\eqref{PFRW} and~\eqref{ferm_proj_FRW}
\begin{align*}
P-P_\text{\tiny{WKB}} &= \mathcal{N} \:k_m - \mathcal{N}_\text{\tiny{WKB}} \:k_\text{\tiny{WKB}} \cr
&= \left( \mathcal{N} - \mathcal{N}_\text{\tiny{WKB}} \right) k_m + \mathcal{N}_\text{\tiny{WKB}} \left( k_m - k_\text{\tiny{WKB}} \right) .
\end{align*}
Applying a test function~$\phi \in C^\infty_0((0, \pi), \C^2)$ and taking the norm,
we can use that~${\mathcal{N}}_\text{\tiny{WKB}}$ has norm at most one to obtain
\beq \label{PmP}
\|(P-P_\text{\tiny{WKB}})(\phi)\|
\leq \|(k_m - k_\text{\tiny{WKB}})(\phi)\| + \|\mathcal{N} - \mathcal{N}_\text{\tiny{WKB}}\|\:
\| k_m(\phi) \| \:.
\eeq

In order to estimate the first summand in~\eqref{PmP}, we first note that, according to~\eqref{kmFRW}
and~\eqref{kmWKB},
\beq \label{kmkWKB}
(k_m - k_\text{\tiny{WKB}})(\phi) = \frac{1}{2\pi} \int_0^\pi (U^{\tau, \tau_0} - U^{\tau, \tau_0}_{\text{\tiny{WKB}}})^* 
\begin{pmatrix}1 & 0 \\ 0 & -1 \end{pmatrix} \phi(\tau) \:R(\tau)\: d\tau\:.
\eeq
Using~\eqref{Wdef} and Lemma~\ref{lemma63}, we get the estimate
\[ \big\| (k_m - k_\text{\tiny{WKB}})(\phi) \big\|
\leq \frac{c \,|\lambda|}{m} \int_0^\pi \|\phi(\tau)\|\, d\tau \:. \]
In the case~$|\lambda| \leq (m R_{\max})^\frac{4}{5}$, we get the inequality
\beq \label{kmWKBes} \big\| (k_m - k_\text{\tiny{WKB}})(\phi) \big\|
\leq c\: (m R_{\max})^{-\frac{1}{5}}\: R_{\max} \int_0^\pi \|\phi(\tau)\|\, d\tau \:.
\eeq
In the remaining case~$|\lambda| > (m R_{\max})^\frac{4}{5}$, we
apply Lemma~\ref{lemmaes3} to~\eqref{kmkWKB}, again giving~\eqref{kmWKBes}.

It remains to estimate the second summand in~\eqref{PmP}. Noting that
\begin{align*}
\|k_m(\phi)\| &\leq \frac1{2\pi} \int_0^\pi 
\| \phi(\tau)\| \:R(\tau)\: d\tau \leq \frac1{2\pi}\: R_{\max} \int_0^\pi \|\phi(\tau)\|\, d\tau \:,
\end{align*}
the proof of the theorem is completed by applying Lemma~\ref{lemmanext} below.
\QED

\begin{Lemma} \label{lemmanext} Under the assumptions of Theorem~\ref{thmPmP},
\[ \|{\mathcal{N}} - {\mathcal{N}}_\text{\tiny{WKB}}\| \leq c\, (m R_{\max})^{-\frac{1}{5}}\:. \]
\end{Lemma}
\Proof Writing the spectral calculus with residues, we have
\[ \mathcal{N} = \frac{1}{2\pi i} \ointctrclockwise_\Gamma \left(\Sig-\zeta\right)^{-1} d\zeta \:, \]
where~$\Gamma$ is a curve in the left half plane enclosing all negative eigenvalues. Similarly,
\[ \mathcal{N}_\text{\tiny{WKB}} = \frac{1}{2\pi i} \ointctrclockwise_{\Gamma_\text{\tiny{WKB}}}
\left(\Sig_\text{\tiny{WKB}}-\zeta\right)^{-1} d\zeta \:. \]
We choose~$\Gamma_\text{\tiny{WKB}}$ as a circle centered at the negative
eigenvalue~$\mu^-_\text{\tiny{WKB}}$ with radius $r=|\mu^-_\text{\tiny{WKB}}|/2$.
Using~\eqref{eigval_S_WKB} together with~\eqref{lames}, we can estimate this eigenvalue by
\begin{align*}
|\mu^-_\text{\tiny{WKB}}| &\geq \int_0^\pi \frac{m R^2}{\sqrt{\lambda^2 + m^2 R^2}}\: d\tau \\
&\geq \int_0^\pi  \frac{m R^2}{\sqrt{K^2 m^2 R_{\max}^2 + m^2 R^2}}\: d\tau
= c R_{\max}\:,
\intertext{where}
c &:= \int_0^\pi  \frac{g^2}{\sqrt{K^2+g^2}}\: d\tau > 0\:.
\end{align*}

According to Proposition~\ref{bound_|S-SWKB|_Lemma}, we can treat the
operator~$\Delta \Sig:=\Sig_\text{\tiny{WKB}}-\Sig$ as a perturbation. More precisely,
the min-max-principle (see for example~\cite{reed+simon4})
yields that the negative eigenvalue of the operator~$\Sig+t \Delta \Sig$, $t \in [0,1]$,
lies inside~$\Gamma$, and that the distance of the eigenvalues of all these operators 
from~$\Gamma$ is at least equal to
\beq \label{ddef}
d := \frac{r}{2} \geq \frac{c R_{\max}}{4}\:.
\eeq
It follows that
\begin{align}
\mathcal{N} - \mathcal{N}_\text{\tiny{WKB}}
&= -\frac{1}{2\pi i} \ointctrclockwise_\Gamma \left[ \left(\Sig-\zeta\right)^{-1} - ( \Sig_\text{\tiny{WKB}} - \zeta )^{-1} \right] d\zeta \cr
&= \frac{1}{2\pi i} \ointctrclockwise_\Gamma \int_0^1 \frac{d}{dt} \left(\Sig+t\Delta \Sig - \zeta  \right)^{-1} dt \,d\zeta \cr
&= -\frac{1}{2\pi i} \ointctrclockwise_\Gamma \int_0^1 \left(\Sig+t\Delta \Sig - \zeta \right)^{-1} \Delta \Sig \left(\Sig+t\Delta \Sig - \zeta \right)^{-1} dt \,d\zeta \:,
\end{align}
where we set~$\Delta \Sig=\Sig_\text{\tiny{WKB}}-\Sig$. Taking the norm and estimating gives
\[ \| \mathcal{N} - \mathcal{N}_\text{\tiny{WKB}} \|
\leq \frac{1}{2\pi} \max_{t \in [0,1]} \ointctrclockwise_\Gamma \left\| (\Sig+t\Delta \Sig - \zeta)^{-1} \right\|^2 \|\Delta \Sig\|\: d|\zeta| 
\leq \frac{r}{d^2}\: \|\Delta \Sig\|\:. \]
Applying~\eqref{ddef} and Proposition~\ref{bound_|S-SWKB|_Lemma} gives the result.
\QED

\section{Discussion of Examples with a Piecewise Constant Scale Function} \label{seccounter}
Qualitatively speaking, the results of Section~\ref{secFRW} show that our definition of the
fermionic projector reduces to the naive notion of the Dirac sea as ``all solutions of negative frequency,''
provided that the metric is nearly constant on the Compton scale.
This raises the question what happens if the metric varies substantially on the Compton scale.
In order to tackle this question, we now analyze the situation for a closed Friedmann-Robertson-Walker
space-time with a piecewise constant scale function. This analysis is also instructive because it will give a connection to the well-known Klein paradox.

We again consider the line element~\eqref{lineelement_FRW}. Again separating the spatial dependence,
the operator~$\Sig$ is given by~\eqref{S_FRW}, where the unitary matrix~$U^{\tau, \tau_0}$
is defined as the solution of the initial value problem~\eqref{evolution_eqn_FRW} and~\eqref{initval}.
In order to get a better geometric understanding of the dynamics, it is useful to decompose 
the matrix in the integrand of~\eqref{S_FRW} in terms of Pauli matrices by setting
\[ v_\alpha(\tau) := \frac{1}{2}\: \Tr \left( \sigma_\alpha \:U^{0, \tau}_m \begin{pmatrix} 1 & 0 \\ 0 & -1 \end{pmatrix} U^{\tau, 0}_m \right) . \]
After cyclically commuting the factors in the trace, we obtain
\beq \label{vwrel}
v_\alpha(\tau) = \la \vec{w}_\alpha(\tau), e_3 \ra \:,
\eeq
where~$\vec{w}_\alpha(\tau)$ is (for any given~$\alpha =1,2,3$) the vector
\beq \label{wdef}
\vec{w}_\alpha(\tau) = \frac{1}{2}\, \Tr \left( \vec{\sigma} \: U^{\tau, 0}_m \sigma_\alpha \:U^{0, \tau}_m  \right) .
\eeq
Taking the $\tau$-derivative and using~\eqref{evolution_eqn_FRW} gives
\[ \partial_\tau \vec{w}_\alpha(\tau) =
\frac{1}{2} \Tr \bigg( \vec{\sigma} \:\Big[ i \frac{\vec{d} \vec{\sigma}}{2},  U^{\tau, 0}_m \sigma_\alpha \:U^{0, \tau}_m \Big] \bigg) 
= \frac{i}{4} \Tr \Big( \big[\vec{\sigma}, \vec{d} \vec{\sigma} \big]  U^{\tau, 0}_m \sigma_\alpha \:U^{0, \tau}_m
\Big)
\:, \]
where the vector~$\vec{d}$ has the components
\beq \label{drot}
\vec{d} = 2 \,(\lambda, 0, -m R)\:.
\eeq
Using the commutation relations of the Pauli matrices, we obtain
\beq \label{bloch}
\partial_\tau \vec{w}_\alpha = \vec{d} \wedge \vec{w}_\alpha\:.
\eeq
Moreover, evaluating~\eqref{wdef} for~$\tau=0$ gives the initial condition
\beq \label{winit}
\vec{w}_\alpha(0) = \vec{e}_\alpha \:.
\eeq
The differential equation~\eqref{bloch} describes a rotation of the vector~$\vec{w}$ around the axis~$\vec{d}$,
which also depends on~$\tau$. This equation can be regarded as the {\em{Bloch representation}}
of the Dirac equation~\eqref{evolution_eqn_FRW} (see the discussion of the Dirac equation in~\cite[Section~2]{bounce}). However, the initial conditions~\eqref{winit} and the connection to the vector~$\vec{v}$ by~\eqref{vwrel}, are specific to the construction of the fermionic projector.

Next, we choose the scale function~$R(\tau)$ to be piecewise constant.
Thus we introduce intermediate points~$\tau_0=0 < \tau_1 < \cdots < \tau_N=\pi$ and set
\[ R(\tau) = \sum\limits_{n=1}^N R_n \: \chi_{[\tau_{n-1}, \tau_n)}(\tau) \]
with parameters~$R_1, \ldots, R_N > 0$.
Then on the subinterval~$[\tau_{n-1}, \tau_n)$, the dynamics of~\eqref{bloch} reduces to
the rotation of the Bloch vector around the fixed rotation axis
\[ \vec{d}_n := 2 \,(\lambda,0, -mR_n) \:. \]
The angular velocity of this rotation is given by~$2 \sqrt{\lambda^2 + m^2 R_n^2}$.
This is the frequency of the so-called Zitterbewegung of the Dirac particle;
it is twice the frequency of the oscillations of the Dirac wave functions.
We denote the number of full rotations of the Bloch vector on the interval~$[\tau_{n-1}, \tau_n)$
by~$p_n$. Then
\[ \tau_n - \tau_{n-1} = \frac{\pi\, p_n}{\sqrt{\lambda^2 + m^2 R_n^2}}\:. \]

If the scale function is constant, we may
decompose the spinors into eigenfunctions of the matrix~$\vec{d} \vec{\sigma}$. This corresponds precisely
to the splitting of the solutions into solutions of positive and negative frequency.
However, this splitting depends on the value of the scale function. In particular,
if~$R$ changes discontinuously, the canonical splitting into positive and negative frequency
solutions gets lost. Nevertheless, the fermionic projector is well-defined.
Let us analyze how this comes about:
According to~\eqref{S_FRW}, we must integrate~$\vec{v}$ over time,
\beq \label{Sint}
\vec{\Sig} = \int_0^\pi \vec{v}(\tau)\, R(\tau)\, d\tau \:.
\eeq
As a consequence, only the ``time average'' of~$\vec{v}$ enters the construction,
but a canonical splitting of the solution space into solutions of positive and negative
frequency is no longer needed.
This time average means that the fermionic
projector will be an ``interpolation'' of the concepts of negative frequency before
and after the step potential (for a related discussion of a scattering process
see~\cite[Section~5]{sea}). This interpolation is performed in such a way
that the construction of the fermionic projector is manifestly covariant and independent
of observers.

The last explanation also applies to Klein's paradox. Namely, in the setting of the classical Klein's paradox
(see for example~\cite[Section~3.3]{bjorken} or~\cite[Section~4.5]{thaller}),
one considers a potential barrier, i.e.\ an electric potential which is time-independent but has a discontinuous
spatial dependence. If the amplitude of this potential exceeds the mass gap,
the frequency of the solutions no longer gives a natural splitting of the solution space
of the Dirac equation into two subspaces. However, this does not cause any problems in
the construction of the fermionic projector, where in analogy to~\eqref{Sint} a space-time
average is taken (cf.~\eqref{Sdef} or Lemma~\ref{sgn_operator_foliation}).

According to~\eqref{bloch} and~\eqref{drot}, the Bloch vector~$\vec{w}$
rotates around a time-dependent rotation axis~$\vec{d}$.
This can lead to bizarre effects when the rotation axis is tilted several times in a fine-tuned way.
In order to illustrate such effects in a simple example, we now construct a
space-time where~$\Sig=0$. In this case, the fermionic projector defined by~\eqref{PFRW}
vanishes identically. By slightly changing the geometry, one can perturb the eigenvalues of~$\Sig$.
The operator~$-\chi_{(-\infty, 0)}(\Sig)$ in~\eqref{PFRW} is certainly not stable
under such perturbations. This shows that in a space-time with~$\Sig=0$, the definition of the
fermionic projector suffers from an instability and thus depends sensitively on the detailed
geometry of space-time. From the physical point of view, this shortcoming does not seem to be of any
significance, because the class of space-times with~$\Sig =0$ seems very special and not realistic.

We now introduce our example in detail. With the scale function, we can adjust the
angle of the rotation axis~$\vec{d}$ to the $z$-axis.
We choose~$R_1$ and~$R_2$ such that this angle equals $10^\circ$ resp.\ $70^\circ$, i.e.
\beq \label{R12spec}
R_1 = \frac{\lambda}{m}\: \cot(10^\circ) \:,\qquad R_2 = \frac{\lambda}{m}\: \cot(70^\circ)\:.
\eeq
Moreover, we always choose the parameters~$p_n$ such that we rotate a half-integer times around~$\vec{d}_n$,
so that the rotation amounts to a reflection at the axis~$\vec{d}_n$.
Composing the reflection at~$\vec{d}_1$ with the reflection at~$\vec{d}_2$ gives rise to
a rotation around the axis~$e_2$ about an angle of~$2 \cdot 60^\circ = 120^\circ$.
Repeating this construction three times gives in total a rotation by~$360^\circ$.
More specifically, for the construction so far, we choose~$N=6$ and
\[ (R_n)_{n=1,\ldots, 6} = (R_1, R_2, R_1, R_2, R_1, R_2)\:,\qquad
(p_n)_{n=1,\ldots, 6} = (5.5,0.5, 5.5,0.5, 5.5,0.5) \]
with~$R_1$ and~$R_2$ as in~\eqref{R12spec}
(the choices~$p_1=5.5$ and~$p_2=0.5$ are arbitrary; other half-integer values would work
just as well). Moreover, for convenience we chose~$\lambda=3/2$ and~$m=1$.
We solve the system~\eqref{evolution_eqn_FRW} and~\eqref{initval} with~$\tau_0=0$.
Decomposing the resulting fermionic
signature operator~$\Sig$, \eqref{S_FRW}, in terms of Pauli matrices~\eqref{Sint},
it follows by symmetry in the~$e_1/e_3$-plane that~$\Sig_1=\Sig_3 = 0$. However, $\Sig_2 \neq 0$.

\begin{figure}
\begin{center}
\includegraphics[width=7cm]{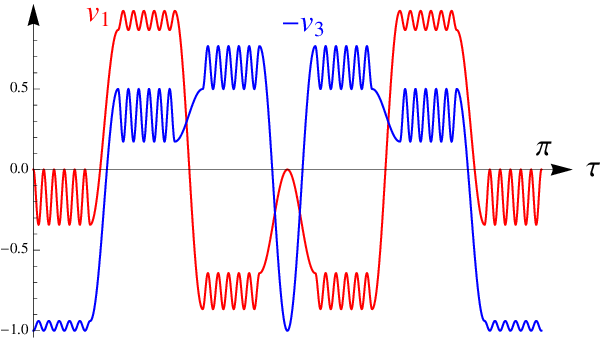}
\includegraphics[width=7cm]{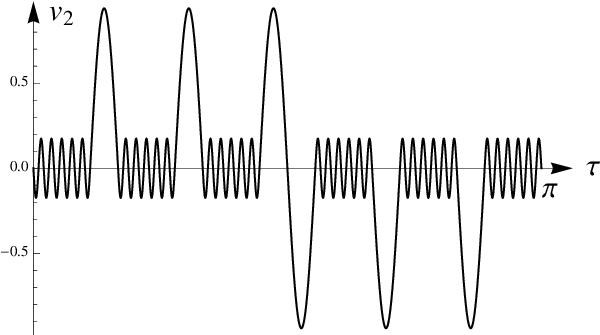}
\end{center} 
\caption{The functions~$v_\alpha(\tau)$.}
\label{figcounter}
\end{figure}

In order to also arrange that~$\Sig_2=0$, we take the above space-time twice, with the opposite time orientation.
Thus we now choose~$N=12$ and
\begin{align*}
(R_n)_{n=1,\ldots, 12} &= \;(R_1, R_2, R_1, R_2, R_1, R_2, \;\;R_2, R_1, R_2, R_1, R_2, R_1) \\
(p_n)_{n=1,\ldots, 12} &= (5.5,0.5, 5.5,0.5, 5.5,0.5, \: 0.5,5.5, 0.5,5.5, 0.5,5.5 ) \:.
\end{align*}
Then the symmetries imply that~$\Sig=0$.
To illustrate the construction, in Figure~\ref{figcounter}
the functions~$v_\alpha(\tau)$ are plotted.

\Thanks{{{\em{Acknowledgments:}}
We would like to thank Olaf M\"uller, Karolin Sporrer, Jan-Hendrik Treude and the referees
for helpful discussions and comments.}

\providecommand{\bysame}{\leavevmode\hbox to3em{\hrulefill}\thinspace}
\providecommand{\MR}{\relax\ifhmode\unskip\space\fi MR }
\providecommand{\MRhref}[2]{%
  \href{http://www.ams.org/mathscinet-getitem?mr=#1}{#2}
}
\providecommand{\href}[2]{#2}

\end{document}